\documentclass[aps,prd,nofootinbib,twocolumn,superscriptaddress,preprintnumbers,balancelastpage,longbibliography]{revtex4-2}
\usepackage{aas_macros}

\usepackage{placeins}
\usepackage{amsmath,amssymb,mathtools,bm}
\usepackage{graphicx, color, hepunits}
\usepackage{gensymb}
\usepackage[dvipsnames]{xcolor}
\usepackage{cancel}
\usepackage[normalem]{ulem}
\usepackage{float}
\usepackage{filecontents}
\usepackage{multirow}
 \usepackage{hyperref} 
\hypersetup{
    colorlinks=true,       
    linkcolor=blue,        
    citecolor=blue,        
    filecolor=magenta,     
    urlcolor=blue          
}
\usepackage[utf8]{inputenc}
\usepackage[english]{babel}
\usepackage{array}

\usepackage{tikz}
\usepackage{tikz-feynman}
\tikzfeynmanset{compat=1.1.0}

\usepackage{booktabs}
\DeclareMathOperator{\Tr}{Tr}

\newcommand{\gagg}{g_{a \gamma \gamma}}

\newcommand{\es}[2] {\begin{equation} \label{#1} \begin{split} #2 \end{split} \end{equation}}

\newcommand{\mycomment}[1]{}

\begin{document}
\title{Blazar Constraints on Axions through New Spectral Modulation Searches in \\ 1ES 1959+650 \& B2 1811+31}

\author{Andrea Giovanni De Marchi}
\affiliation{Dipartimento di Fisica e Astronomia, Universit\`a di Bologna, Bologna 40126, Italy}
\affiliation{INFN, Sezione di Bologna, Bologna 40127, Italy}

\author{Orion Ning}
\affiliation{The Leinweber Institute for Theoretical Physics, University of California, Berkeley, CA 94720, USA}
\affiliation{Theoretical Physics Group, Lawrence Berkeley National Laboratory, Berkeley, CA 94720, U.S.A.}

\author{Tianzhuo Xiao}
\affiliation{The Leinweber Institute for Theoretical Physics, University of California, Berkeley, CA 94720, USA}
\affiliation{Theoretical Physics Group, Lawrence Berkeley National Laboratory, Berkeley, CA 94720, U.S.A.}

\date{\today}

\begin{abstract}
Blazars are unique astrophysical environments whose high-energy $\gamma$-ray spectra are susceptible to modulations in the presence of ultralight axions. We search for these modulations, induced by axion-photon mixing, in Fermi-LAT spectral data of previously unexplored blazar targets, focusing in particular on blazars 1ES 1959+650 and B2 1811+31, whose flare states provide a clean testbed for axion activity. In both cases, we find no evidence for axions, and set exclusion regions on the axion-photon coupling for masses between $10^{-9}$ eV $\lesssim$ $m_a$ $\lesssim$ $10^{-8}$ eV, with sensitivities typically reaching $g_{a \gamma \gamma} \sim 10^{-11} - 10^{-10}$ GeV$^{-1}$ depending on the assumed blazar modeling choices. We examine the broad impact of modeling uncertainties, finding that the resulting constraints can vary substantially across plausible configurations. We discuss the implications of these systematic effects and their relevance for similar blazar-like searches in the future. 
\end{abstract}
\maketitle



\section{Introduction}

The axion is a strongly motivated particle candidate that solves multiple problems in fundamental physics. Introduced to solve the strong CP problem of quantum chromodynamics (QCD), the axion is able to both explain a vanishing neutron electric dipole moment, as well as serve as a natural dark matter candidate, potentially produced non-thermally in the early Universe~\cite{Peccei:1977hh, Peccei:1977ur, Weinberg:1977ma, Wilczek:1977pj, Preskill:1982cy, Abbott:1982af, Dine:1982ah}. While field theory axions act as the pseudo-Nambu Goldstone boson of the $U(1)_{\rm PQ}$ symmetry after spontaneous symmetry breaking, the additional fact that they can be produced from string theory compactifications from higher-dimensional constructions provides yet another motivation for axions across a wide range of masses (\textit{i.e.} the string `axiverse')~\cite{Witten:1984dg, Choi:1985je, Barr:1985hk, Svrcek:2006yi, Arvanitaki:2009fg, Demirtas:2018akl, Halverson:2019cmy, Mehta:2021pwf, Gendler:2023kjt,Benabou:2025kgx}.

Axions, which we refer to in this work as both the QCD axion solving the strong CP problem as well as generic axion-like particles (ALPs), have a rich phenomenology leading to a variety of search strategies. The axion effective field theory (EFT) contains the generic term $\mathcal{L} \supset -\gagg a F_{\mu \nu}\tilde{F}^{\mu \nu}/4$, allowing axions to couple to Quantum Electrodynamics (QED) and the photon, with $\gagg$ the dimensionful axion-to-photon coupling, $a$ the axion field, and $F_{\mu \nu}$ the electromagnetic field strength tensor. While searches, including the one in this work, focus on the axion-to-photon coupling $\gagg$, other terms in the full axion EFT, such as couplings to fermions, can be related to the strength of $\gagg$ depending on the UV completion of the axion (see, \textit{e.g.},~\cite{Srednicki:1985xd,Chang:1993gm,Dine:1981rt,Zhitnitsky:1980tq,Cicoli:2012sz,Choi:2021kuy,Choi:2024ome,Reece:2024wrn}). The advantage of axion-to-photon coupling searches, however, is the ease of observing direct photon signals for signs of new physics, especially in astrophysical phenomena. Various searches probing $\gagg$ in astrophysics often involve direct axion production from, \textit{e.g.} stars, which can then convert to observable photons in external magnetic fields of the star, the host galaxy, or even terrestrial laboratory experiments~\cite{Anastassopoulos:2017ftl,Dessert:2020lil,Xiao:2020pra,Ning:2024eky,Manzari:2024jns,Ning:2025fqd}. These types of searches, often called astrophysical light-shining-through-walls, remain among the most popular direct ways to find axions (for a review see, \textit{e.g.},~\cite{Caputo:2024oqc, Safdi:2022xkm}).

On the other hand, another interesting probe for evidence of axions that rely on $\gagg$ involves the process of spectral modulation. In this scenario, instead of searching for a direct axion-induced photon signal, one can leverage the fact that the axion-photon coupling allows for photons ejected from astrophysical sources to experience axion conversion and photon re-conversion, effectively inducing \textit{modulations} in the resulting observed photon spectra. These types of searches have been proven to be powerful, setting stringent constraints on the ultralight axion parameter space which we focus on in this work~\cite{TheFermi-LAT:2016zue, Li:2020pcn, Li:2021gxs, Li:2024zst, Zhou:2025txk} (see also X-ray analogs~\cite{Reynolds:2019uqt, Reynes:2021bpe}). As we will show, for ultralight axion masses $m_a \sim 10^{-9} - 10^{-7}$ eV, this effect is particularly magnified in high-energy GeV-TeV environments, such as the $\gamma$-rays produced in blazars. Blazars, which are a special type of active galactic nuclei (AGN), have powerful high-energy jets of particles formed from accretion on supermassive black holes, and are pointed toward Earth. They provide a unique testbed for axion physics through spectral modulation, as they emit over many decades in energy, have substantial magnetic fields which facilitate the conversion, and possess intrinsic astrophysical spectra which are generally smooth and predictable.  

The usage of blazars for axion spectral modulations have been appreciated with several existing works using several blazar targets through a variety of GeV-TeV observations. In particular, stringent constraints have been obtained with multiple analyses of blazar Markarian 421 (Mrk 421)~\cite{Li:2020pcn, Li:2021gxs, Li:2024zst}, one of the brightest TeV sources that also undergoes regular outbursts in the $\gamma$-ray band. In these analyses, the non-observations of axions were deduced through Fermi-LAT, ARGO-YBJ, and MAGIC observations of the various steady and flaring phases of Mrk 421. This is supplemented by a similar work analyzing Mrk 501 with Fermi-LAT and MAGIC, while also exploring various modeling assumptions sourcing the magnetic field of the blazar~\cite{Zhou:2025txk}. Other recent, similar analyses that also search for axion-induced modulations from $\gamma$-ray flares (not necessarily from blazars) include using Fermi-LAT and VERITAS observations of NGC 1275 in the Perseus cluster to constrain $\gagg$~\cite{TheFermi-LAT:2016zue, VERITAS:2025ett}.

\begin{figure}[!htb]
\centering
\includegraphics[width=0.49\textwidth]{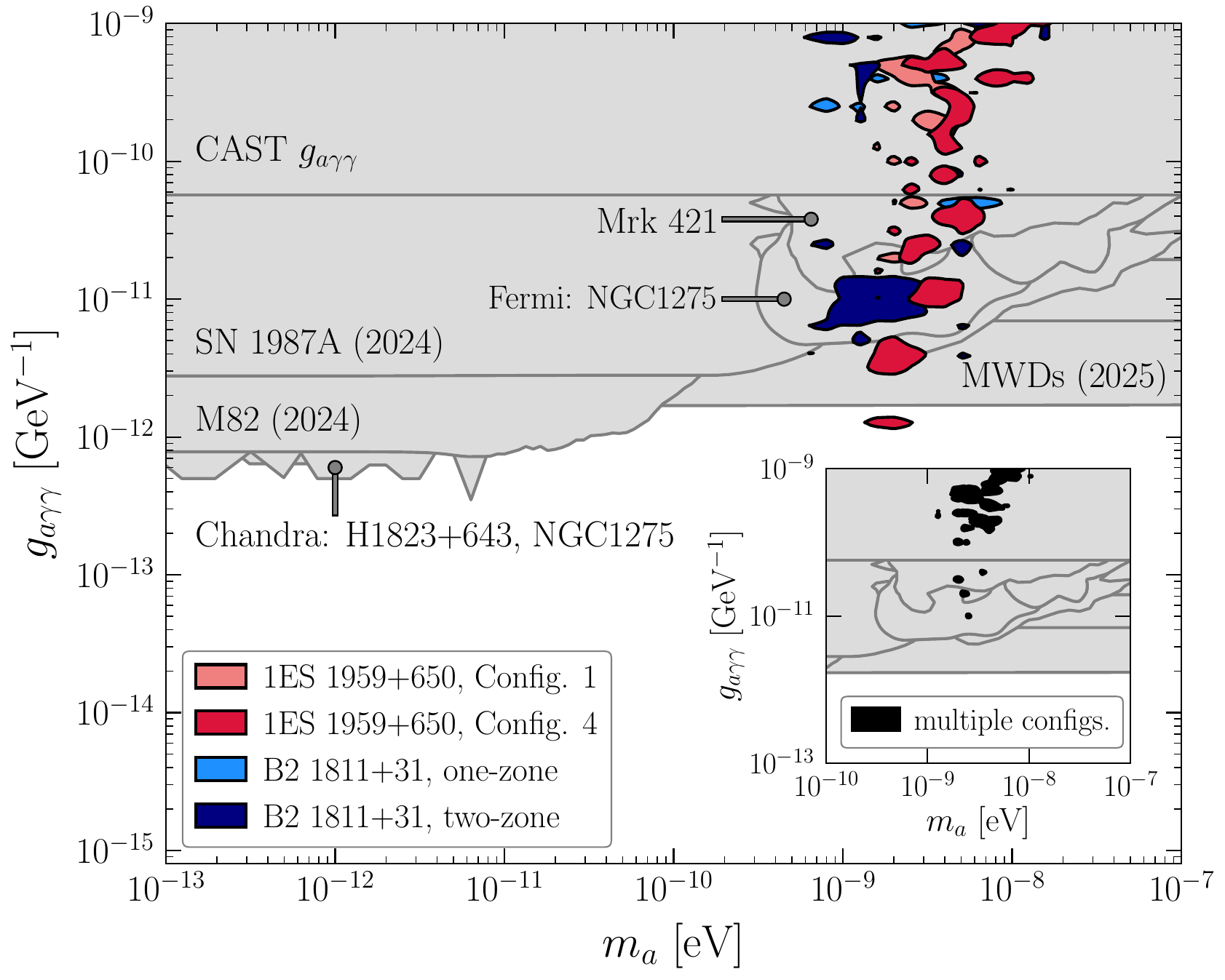}
\vspace{-0.4cm}
\caption{The configuration-dependent 95\% exclusion regions on $\gagg$ obtained in this work from the non-observation of axion-induced spectral modulations in the $\gamma$-ray spectra of our blazar targets, 1ES 1959+650 and B2 1811+31. We illustrate representative constraints from several fiducial astrophysical configurations: Configurations 1 and 4 for 1ES 1959+650, corresponding to two plausible choices of magnetic field strength and emission region size, and the one-zone and two-zone emission models for B2 1811+31 (note that the one-zone model is disfavored through complementary observations, see main text). The inset highlights the regions simultaneously excluded by multiple configurations for each blazar (among all configurations considered for each), indicating parameter space where constraints would likely be more robust against modeling uncertainties. We compare to existing constraints in gray~\cite{AxionLimits}.}
\label{fig:exclusions}
\end{figure}

In our work we adopt a similar framework and search for axion-induced spectral modulations in new blazar targets previously unexplored in the context of axions. We study the blazars 1ES 1959+650 and B2 1811+31, which are motivated by their relatively smooth spectra, substantial magnetic fields, and large emission region sizes. These properties make them potentially promising environments for spectral modulation searches. We highlight the ability of these targets to test axion parameter space in the $\gagg-m_a$ plane, and examine how the resulting constraints depend on the underlying systematic uncertainties. We illustrate the parameter space probed in this work for several configurations of our two blazars in comparison with existing constraints in Fig.~\ref{fig:exclusions}, with the inset showing more restrictive regions simultaneously excluded by multiple configurations for each blazar, providing a more realistic indication of excluded parameter space when taking into account the effects of systematic modeling uncertainties.



This work is organized as follows. In Sec.~\ref{sec:axion_spectral_modulation} we detail the formalism of axion-induced modulation in astrophysical spectra and the application toward blazars. In Sec.~\ref{sec:blazar_env} we transition to describing the blazar environment and some of the various modeling assumptions that underlie these types of analyses. In Sec.~\ref{sec:targets_data} we describe our blazar targets of interest in this work, their data, and the data analysis procedure leading to our final constraints on the axion-photon coupling $\gagg$ which we discuss and compare across modeling configurations in Sec.~\ref{sec:results_systematics}. In Sec.~\ref{sec:discussion} we discuss the implications of our findings as well as future directions. 


\section{Axion-Induced Spectral Modulations}
\label{sec:axion_spectral_modulation}

Photons propagating through an astrophysical magnetic field $\mathbf{B}$ can oscillate into axions and back, producing characteristic modulations in their observed energy spectra. As mentioned earlier, this phenomenon arises from the following interaction term in the axion EFT Lagrangian between the axion $a$ and QED,

\begin{equation}
    \mathcal{L} \supset -\frac{1}{4} g_{a\gamma\gamma} a F_{\mu\nu} \tilde{F}^{\mu\nu} = \gagg a \mathbf{E} \cdot \mathbf{B} \,.
\end{equation}
The form of this term implies that the axion mixes with the photon mode $A_\parallel$  of polarization parallel to the transverse component of the external magnetic field $\mathbf{B_T}$, forming a three-state vector $\Psi=(A_\perp, A_\parallel, a)$. 

Faraday rotation is negligible at the energies relevant to this work, and following the discussion in Ref.~\cite{PhysRevD.37.1237}, the entire effect of mixing can be encoded into a matrix $\mathcal{M}$ describing the equations of motion for this axion-photon system

\begin{equation}
i\frac{d}{ds}
\begin{pmatrix}
A_\perp \\
A_\parallel \\
a
\end{pmatrix}
=
\begin{pmatrix}
\Delta_{\parallel} & 0 & 0 \\
0 & \Delta_{\perp} & \Delta_{a\gamma} \\
0 & \Delta_{a\gamma} & \Delta_a
\end{pmatrix}
\begin{pmatrix}
A_\perp \\
A_\parallel \\
a
\end{pmatrix},
\end{equation}
where $s$ is the distance along the direction of propagation, \( \Delta_{a\gamma}= g_{a\gamma\gamma} B_T/2 \), \( \Delta_a=-m_a^2/(2E) \), and \( \Delta_{\parallel,\perp} \) include plasma, cosmic microwave background (CMB) and QED vacuum contributions~\cite{PhysRevD.37.1237}. Physically, $\Delta_{\parallel,\perp}$ encode polarization-dependent refractive effects that modify the photon dispersion relation during propagation.

The main observable of this process is the effect of the photon survival probability on the spectra, $P_{\gamma\gamma}=1-P_{a\gamma}$, whose counterpart is the photon conversion probability $P_{a\gamma}$. The approximate analytic form of $P_{a\gamma}$ in the presence of a simple, constant transverse magnetic field in vacuum is well-known~\cite{Sikivie:1983ip,Mirizzi:2007hr}:
\begin{equation}
    P_{a \gamma}(E) \simeq \left( \frac{\gagg B_T}{\Delta k} \right)^2 \sin^2\left( \frac{\Delta k L}{2}\right) \,,
\end{equation}
where $\Delta k \simeq  m_a^2/(2E)$ represents the momentum mismatch between the axion and photon and $L$ is the propagation length scale of the system. The quadratic dependence of the survival and conversion probability on $\gagg$ and the transverse magnetic field, as well as the dependence through the length $L$, provide clear guidance in the selection of blazar targets, favoring sources with strong transverse magnetic fields and large coherence lengths where spectral modulations would be maximized.

\begin{figure*}[!htb]
\centering
\includegraphics[width=0.49\textwidth]{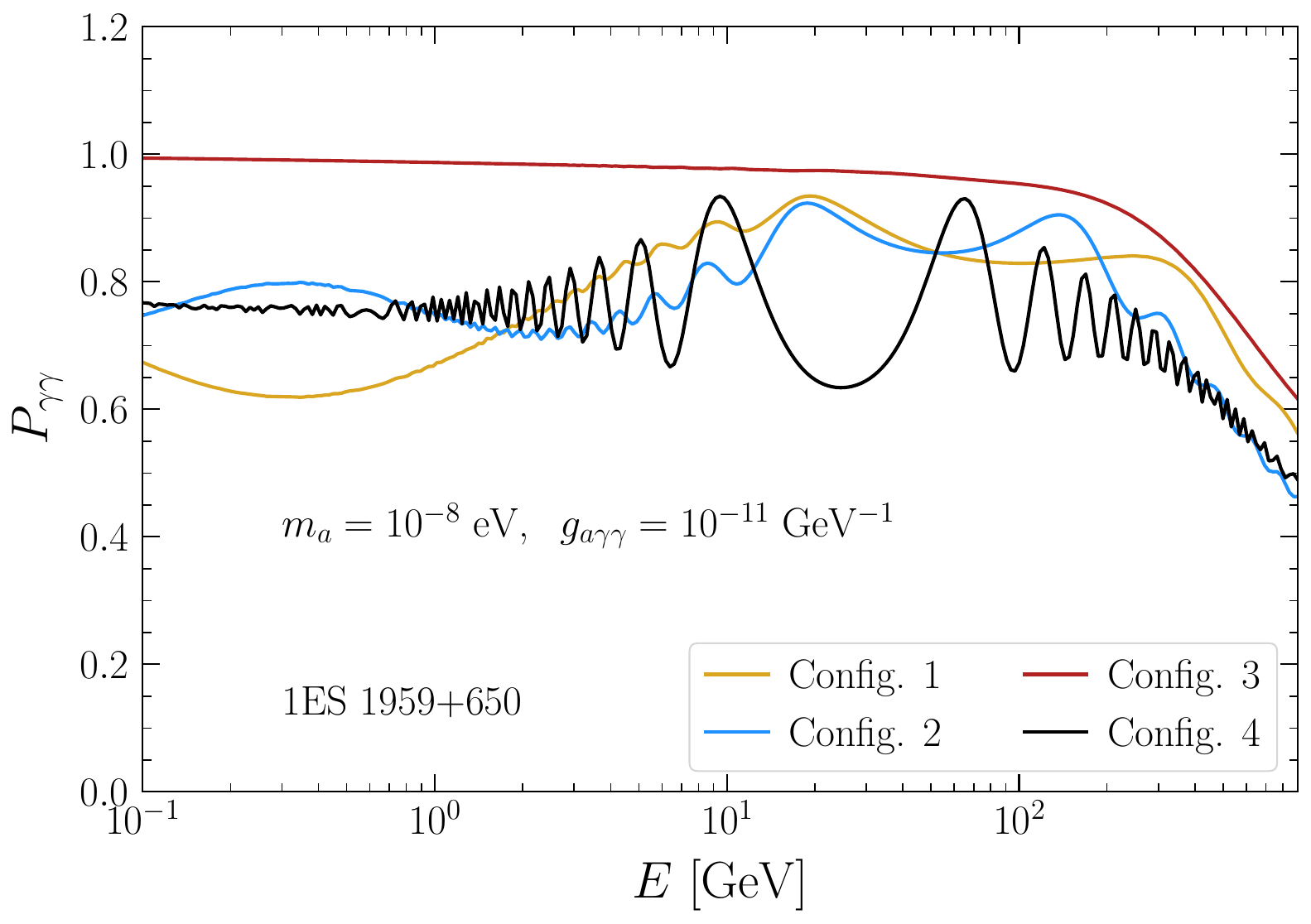}
\includegraphics[width=0.49\textwidth]{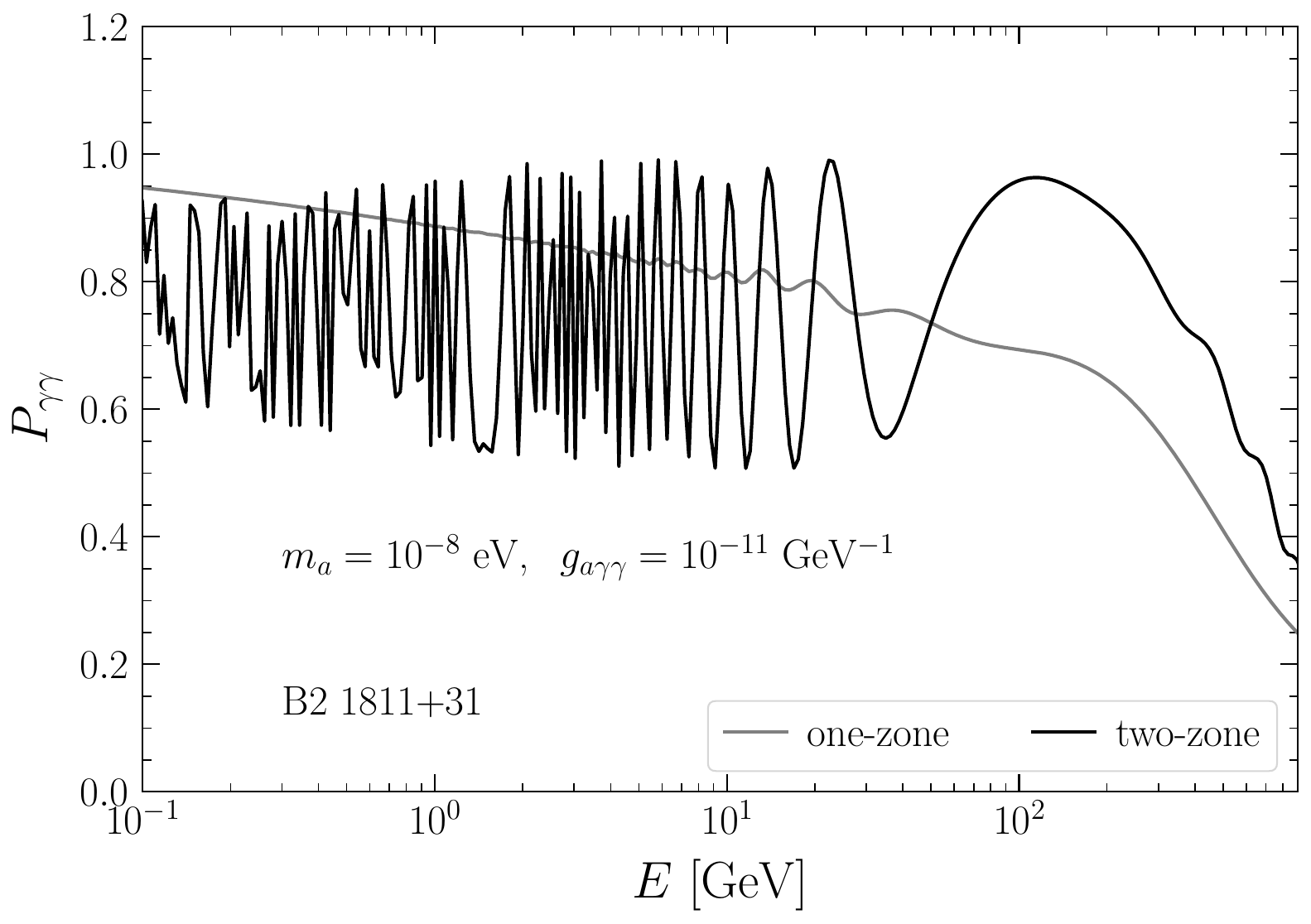}
\vspace{-0.4cm}
\caption{(Left) The survival probability at the given axion parameters $m_a$ and $\gagg$ as a function of energy, for our four configurations of blazar target 1ES 1959+650 (see Table~\ref{tab:blazar_targets}). (Right) The same but for our one/two-zone models of blazar target B2 1811+31.  }
\label{fig:prob_survival}
\end{figure*}

For a real astrophysical environment, we apply numerical methods to calculate $P_{\gamma \gamma}$, in practice utilizing the \texttt{gammaALPs} code package~\cite{Meyer:2021} throughout our work. These calculations for $P_{\gamma \gamma}$ operate under the density matrix formalism $\rho=\Psi\otimes\Psi$~\cite{Galanti:2018nvl}, where the photon survival probability can be expressed as

\begin{equation}
    P_{\gamma\gamma}=\Tr \, [\rho_{f,p} T(s_f,s_i) \rho_i T^\dagger(s_f,s_i)] \,,
\end{equation}
with $T(s_f,s_i)$ the transfer matrix describing the spatial propagation of the photon-axion system between positions $s_i$ and $s_f$. In other words, the density matrix evolves with distance according to $\rho(s_f)=T(s_f,s_i)\rho_iT^{\dagger}(s_f,s_i)$. Here, $\rho_i$ denotes the initial beam density matrix, and $\rho_{f,p}$ are projection operators onto the final photon polarization states $p\in \{\parallel, \perp\}$. Consequently, these density matrices become


\begin{equation}
    \rho_i
    =
    \begin{pmatrix}
        \frac{1}{2} & 0 & 0 \\
        0 & \frac{1}{2} & 0 \\
        0 & 0 & 0 \\        
    \end{pmatrix}
    \hspace{0.4cm} {\rm and} \hspace{0.4cm}
    \rho_{f,p}
    =
    \begin{pmatrix}
    \delta_{\parallel,p} & 0 & 0 \\
    0 & \delta_{\perp,p} & 0 \\
    0 & 0 & 0 \\
    
    \end{pmatrix} \,,
\end{equation}
where we assume that the $\gamma$-ray beam is initially unpolarized and contains no axion component. The total photon survival probability is obtained by summing over the two final polarization states. In Fig.~\ref{fig:prob_survival} we illustrate the calculated survival probabilities over a broad energy range for our two blazar targets and their respective configuration choices, which we expand on in Sec.~\ref{sec:blazar_env}.


The $\gamma$-ray photons additionally experience absorption by extragalactic background light (EBL), which exponentially suppresses the flux above tens of GeV, and is a known source of attenuation~\cite{Cooray:2016jrk}. The resulting optical depth, $\tau_{\rm EBL}(E,z)$, can be computed following the standard formalism in Ref.~\cite{Gould:1967zza}, and enters the predicted flux through the multiplicative factor $e^{-\tau_{\rm EBL}}$. Thus, the combination of axion-photon mixing and EBL contribute to the attenuation of the observed blazar spectrum. In this work we adopt the semi-empirical canonical model in Ref.~\cite{Primack:2011ny} for $\tau_{\rm EBL}$, which empirically reconstructs the EBL from galaxy SEDs and luminosity functions, consistent with Fermi-LAT~\cite{Fermi-LAT:2018lqt} and HESS~\cite{HESS:2017vis}.


\section{The Blazar Environment and Targets}
\label{sec:blazar_env}
In this section we give further descriptions of salient blazar characteristics and the physics of the various sources of $\gamma$-ray emission coming from blazars.
As a brief review, active galactic nuclei (AGN) are bright compact regions found in the cores of a fraction of galaxies, often outshining their host, powered by accreting supermassive black holes. A subset of AGN launch relativistic jets of plasma perpendicular to their accretion disk, and are correspondingly named \textit{jetted AGN}. When the jet forms an angle $\theta_{\text{los}} \lesssim 10\degree-15\degree$ with respect to our line of sight, these objects are classified as blazars. They are extremely bright $\gamma$-ray emitters, accounting for the majority of resolved extragalactic sources detected by Fermi-LAT \cite{Fermi-LAT:2022byn, Ballet:2023qzs}, and are expected to contribute significantly to the unresolved extragalactic $\gamma$-ray background \cite{Fermi-LAT:2018udj, Korsmeier:2022cwp}. Their extreme emissions are further enhanced by the relativistic beaming effect. 

The blazar EM spectrum is typically characterized by two peaks, one in the infrared/soft X-ray band and a second one in the $\gamma$-ray range, both of non-thermal origin. Explaining the origin of this peculiar emission spectrum is an ongoing effort, but models generically include one or more emitting regions represented by spherical ‘blobs’ of plasma streaming along the jet axis at relativistic speed. Within the blob, a population of relativistic particles emits non-thermally via interactions with the highly magnetized environment and ambient photons. In this picture, the low-energy peak is the result of electron synchrotron emission in the intense magnetic fields of the blob, which results in polarized emissions that can be used to probe the structure of the magnetic field~\cite{galaxies7040085}. 

The high-energy peak, on the other hand, can be explained by a variety of mechanisms classified according to the particles responsible for this emission. In \textit{hadronic models}, the peak is the result of proton synchrotron emission, while in \textit{leptonic models} it is due to electron inverse Compton (EIC) scattering, either onto ambient photons, seeded by the accretion disk, dusty torus and broad-line-regions, or onto their own low-energy synchrotron photons, a process named Synchrotron self-Compton (SSC) (for a review of radiative processes in blazar jets, see, \textit{e.g.},~\cite{Cerruti:2020lfj} and references therein). Purely hadronic models are generally disfavoured for two main reasons: First, to correctly reproduce the observed SED, these models require very intense magnetic fields and proton power, disfavoring them on energetic grounds~\cite{Cerruti:2020lfj}. Second, the detection of high-energy neutrinos from this class of objects by the IceCube experiment, from the source TXS 0506+056~\cite{IceCube:2018dnn, IceCube:2018cha} and, subsequently, others~\cite{Kadler:2016ygj, Fermi-LAT:2019hte, Rodrigues:2020fbu, Sahakyan:2022nbz, Oikonomou:2021akf, Giommi:2021bar}, confirm that although hadronic processes must take place in blazars, purely hadronic models predict neutrino fluxes that severely undershoot the observed emissions. This has led to the development of mixed scenarios, named \textit{lepto-hadronic models}~\cite{Cerruti:2018tmc, 10.1093/mnrasl/slaa188, Keivani:2018rnh, Gao:2018mnu}, where the gamma-ray photons are dominantly sourced by EIC but they receive a contribution from the decays of $\pi^0$s produced via $p-\gamma$ photo-pion processes. These same processes produce charged pions $\pi^\pm$ whose decays source the observed neutrinos. This class of solutions often requires super-Eddington jet luminosities, which still potentially disfavors them on energetic grounds, but they predict much higher neutrino fluxes. It is worth noting here that the Eddington luminosity is a rough estimate of the maximum radiative power of an accreting object rather than a strict upper limit, and therefore super-Eddington luminosities are not necessarily unphysical. Furthermore, more complex lepto-hadronic models that take into account the evolution of the relevant physical quantities within the vicinity of the SMBH can relieve this tension, requiring sub-Eddington jet power~\cite{Rodrigues:2025cpm}.

These models are further classified depending on the number of emitting regions required to reproduce the observed spectra, with \textit{single-zone models} being the minimal constructions and \textit{multi-zone models} allowing for more flexibility, at the cost of an increased number of parameters. For example, a multi-zone model allows one to separate the sites of low and high-energy emissions, or separate regions for neutrino and $\gamma$-ray production. Regardless of the details of these models, for our purposes the relevant pieces of information are the presence of intense and ordered magnetic fields, well described by a toroidal (orthogonal to the jet’s axis) structure and featuring a $B(r) \propto r^{-1}$ radial dependence, as well as an intrinsic $\gamma$-ray spectrum that can be described through a number of functional forms. In the following, we will describe the emission in terms of a canonical super exponential cutoff power-law (SEPL) model, \textit{i.e.},
\begin{equation}
\begin{alignedat}{1}
\phi_{\rm int}(E) &= \phi_0\left(\frac{E}{\mathrm{GeV}}\right)^{-\alpha}\exp \left(-(E/E_\text{cut})^\beta\right) 
\end{alignedat}
\end{equation}
where $(\phi_0,\alpha,\beta,E_\text{cut})$ are treated as nuisance parameters when fitting our data as described in Sec.~\ref{sec:targets_data}. Note that we explore the alternate log-parabola model in App.~\ref{app:spectra_modeling}, which also results in similar constraints for both our blazars.


As detailed in Sec.~\ref{sec:axion_spectral_modulation}, the intrinsic spectrum is then modified in the jet’s environment and during propagation to Earth by interactions with the extragalactic background light, as well as by photon-axion oscillations in the highly magnetized blazar environment. We remark here that in this context, and in light of the above discussion about blazar properties, the most important blazar parameters relevant to this search would be the magnetic field magnitude $B$ and blob size $R_b$, which then represent the most consequential sources of systematic uncertainties. These quantities are often determined by what type of particle emission model (\textit{i.e.} hadronic, leptonic, lepto-hadronic) a particular analysis assumes.

We will consider two blazars as targets for this search, namely 1ES 1959+650 and B2 1811+31, 
whose parameter configurations are listed in Table~\ref{tab:blazar_targets}. As alluded to in the introduction, these targets are partially motivated by their clean spectra in their flare states which can more easily distinguish modulation signals, but importantly these targets are notable for their relatively strong magnetic fields and blob sizes, as deduced from a multitude of observations. As a measure of uncertainty in the determination of these parameters, in the rest of this section we discuss and state the benchmark values we use to model these blazar targets.

In the case of 1ES 1959+650, we choose four representative parameter configurations for the blazar magnetic field $B$ and blob size $R_b$, all within the framework of a single emission region. These parameters (enumerated in Table~\ref{tab:blazar_targets}) were chosen from a variety of independent observations and analyses of 1ES 1959+650 over various timescales. Importantly, for three of the configurations, we adopt values for $B$ and $R_b$ self-consistent within each work, and for the final configuration we perform a separate light curve analysis at the level of raw data to justify a higher value of $R_b$ compatible with other past observations. Configuration 1~\cite{Wani:2023xbo} derives $B = 0.64$ G and $R_b = 6.56\times10^{15}$ cm. The size of the emission region is derived from variability timescale measurements using X-ray light curve observations in 2018-2020, and the intensity of the magnetic field is inferred through cross-correlating variations in flux across different photon energy bins. This inference arises from the fact that the time lag from one bin to the other is due to the energy dependence of the cooling and acceleration timescales for the electrons in the emission region, which in turn depends on the magnetic field. 
Older X-ray and optical data~\cite{2003A&A...412..711T} observed 1ES 1959+650 in a flare state in 2001, suggesting Configuration 2 with $B = 0.90$ G and $R_b = 9.0 \times10^{15}$ cm. Configuration 3 adopts a conservative choice of parameters ($B = 0.25$ G, $R_b = 1.0 \times 10^{15}$ cm) from the SSC modeling of MAGIC observations~\cite{MAGIC:2020cmk} of flares in 2016, which is partially degenerate with hadronic models with much higher magnetic fields and hence makes the inference of these quantities difficult. Finally, in light of finding larger blob sizes inferred in Ref.~\cite{Patel:2017mrv}, we opt to undertake our own analysis of the flares used directly in our analysis; through $\gamma$-ray light curve analyses we can derive estimates of the blob size $R_b$ through variability timescales $\tau_{\rm var}$ similar to the methodology in Ref.~\cite{Wani:2023xbo}, which bounds $R_b \leq \tau_{\rm var} \, \delta \, c / (1+z)$. Finding consistently long timescales in the actual raw data, we use our light curve analysis to justify and adopt the value $R_b \sim 7.8 \times 10^{16}$ cm from Ref.~\cite{Patel:2017mrv}, with $\tau_{\rm var}$ still consistent with a magnetic field $B = 0.64$ G. These two parameters inform our Configuration 4. We detail our light curve variability analysis in App.~\ref{app:lc_analysis}, where we derive $R_b$ to a similar order as Ref.~\cite{Patel:2017mrv}.

In the case of B2 1811+31, which is less studied, we begin by first considering the emission region as either described by a one-zone and two-zone SSC model, as discussed in Ref.~\cite{Abe:2025ste}. While the one-zone model containing one spherical emission `blob' region can broadly account for the broadband emission, it has known limitations in describing flux in other bands as well as the electron distribution shape. Thus, a two-zone model is proposed in Ref.~\cite{Abe:2025ste} which posits, in addition to the small and energetic `blob,' an extended lower energy `core' region which would dominate optical and UV emission while still contributing to the $\gamma$-ray spectra. We examine both of these scenarios, which call for two different parameter configurations as specified in Ref.~\cite{Abe:2025ste}. Most importantly, the one-zone model suggests $B = 0.13$ G and $R_b = 1.6 \times 10^{16}$ cm, while the two-zone model describes a blob zone of $B = 0.38$ G and $R_b = 4.0 \times 10^{15}$ cm and a core zone of $B = 0.17$ G and $R_b = 2.1 \times 10^{17}$ cm (as listed in Table~\ref{tab:blazar_targets}). 

We will find that the two-zone model can probe significantly more axion parameter space. However, we note an assumption we make in this formalism: We have included both emitting regions sequentially and computed a common survival probability for all photons emitted by the source, which roughly estimates the effect on the entire spectrum. However, the $\gamma$-rays produced in the blazar receive contributions from both emitting regions, which would technically suggest two separate SEPL spectra and separate survival probabilities for each emitting region, with the resulting fluxes added. This would double the number of nuisance parameters and axion conversion probabilities of which to keep track. We have verified, however, that with this more rigorous methodology, the overall resulting bounds on axion parameters would only incur minor changes. Thus, for computational reasons, we operate under the common survival probability and spectral fitting for both regions.

\section{Data Analysis}
\label{sec:targets_data}


We begin by processing the Fermi-LAT data for both blazars. We use the Pass 8 photon data set and use the instrument response functions (IRFs) P8R3$\_$SOURCE$\_$V3 (\textit{evclass=128, evtype=3}), which corresponds to the standard SOURCE class for spectral studies. The data were filtered using the standard quality selections (\textit{DATA$\_$QUAL==1 \&\& LAT$\_$CONFIG==1}). The region of interest (ROI) is selected as $2^\circ$ centered at each source. We included all 4FGL sources within $10^\circ$ as background sources in the likelihood model to avoid biases from nearby source contamination. The galactic and isotropic diffuse emission were modelled using the standard template files \texttt{gll\_iem\_v07.fits} and \texttt{iso\_P8R3\_SOURCE\_V3\_v1.txt}, respectively. The SED-level data points of the target source are extracted in this way using \texttt{Fermitools} and \texttt{Fermipy}. The SED binning is chosen to balance spectral resolution and detection significance (TS $\geq9$) per bin. 


We consider representative phases for each source:
\begin{itemize}
    \item 1ES 1959+650: August 2015 (MJD 57237) to August 2016 (MJD 57601)
    \item B2 1811+31: April 2020 (MJD 58940) to December 2020 (MJD 59190)
\end{itemize}

These selected phases correspond to high-activity periods of each blazar, which are reported in past observations \cite{2024GCN.35746....1B, Abe:2025ste, HESS:2024lbn}. These choices do not exclude the possibility that other phases could also provide constraining results, which we discuss further in App.~\ref{app:phase_sel}.


\setlength{\tabcolsep}{8pt}
\begin{table*}[t]
\centering
\begin{tabular}{ccccccc}
\hline
\multicolumn{7}{c}{Blazar Targets} \\ 
\midrule
Target Name & $z$ & $\delta_D$ & $\theta$ [$^{\circ}$] & $B_0$ [G] & $R_b$ [cm] & Notes \\ 
\midrule
\midrule
1ES 1959+650 & 0.047 & 15.1 & 3.0 & 0.64 & $6.56 \times 10^{15}$ &  Config. 1\\ 
 &  &  &  & 0.90 & $9.0 \times 10^{15}$ &  Config. 2\\ 
 &  &  &  & 0.25 & $1.0 \times 10^{15}$ &  Config. 3\\ 
 &  &  &  & 0.64 & $7.8 \times 10^{16}$ &  Config. 4\\ 
\hline
B2 1811+31 & 0.117 & 10.0  & 5.7 & 0.13 & $1.6 \times 10^{16}$ & one-zone \\ 
 & &  & & 0.38 & $4.0 \times 10^{15}$ & two-zone (blob) \\ 
 & &  & & 0.17 & $2.1 \times 10^{17}$ & two-zone (core) \\ 
\bottomrule
\end{tabular}
\caption{List of the blazar targets considered in this work, along with their properties for each of the fiducial configurations we consider.}
\label{tab:blazar_targets}
\end{table*}

Given our blazar target data, and our axion signal modulation as described in Sec.~\ref{sec:axion_spectral_modulation}, we can derive the expected theoretical modulated $\gamma$-ray spectra as 

\begin{equation}
    \Phi_{\rm th}(E,\, m_a, \, \gagg, \boldsymbol{\theta}) = P_{\gamma \gamma}(E, \, m_a, \, \gagg) \times \phi(E, \, \boldsymbol{\theta})
\end{equation}
where here $\phi$ is the intrinsic spectra $\phi_{\rm int}$ of the blazar, dependent on modeling parameters $\boldsymbol{\theta}$, after EBL attenuation (\textit{i.e.} $\phi = \phi_{\rm int} e^{-\tau_{\rm EBL}}$), and $P_{\gamma \gamma}$ is the survival probability of the photon in the axion environment, as defined in Sec. \ref{sec:axion_spectral_modulation}. The resulting modulated spectra $\Phi_{\rm th}$ is then a function of energy $E$, the axion parameters $m_a, \, \gagg$, and the intrinsic model parameters $\boldsymbol{\theta}$.

This signal model is then forward-modeled through the detector to capture the energy dispersion arising from the finite energy resolution of the detector. For Fermi-LAT, the energy dispersion function can be well approximated by a Gaussian $G(E,\sigma_E)$ of fractional width $\sigma_E/E$~\cite{2009ApJ...697.1071A} across the energy range of focus. The theoretical spectrum $\Phi_{\rm th}$ is therefore convolved with this
Gaussian response $G$ to give the energy-smeared flux averaged over the $i$th reconstructed energy bin [$E_i$, $E_{i+1}$]

\begin{equation}
    \Phi_{\text{sig},i}=\frac{\int^{E_{i+1}}_{E_i}dE\int_0^\infty G(E',E,\sigma_E)\Phi_{\rm th}(E')dE'}{E_{i+1}-E_i} \,.
\end{equation}

\begin{figure*}[!htb]
\centering
\includegraphics[width=0.49\textwidth]{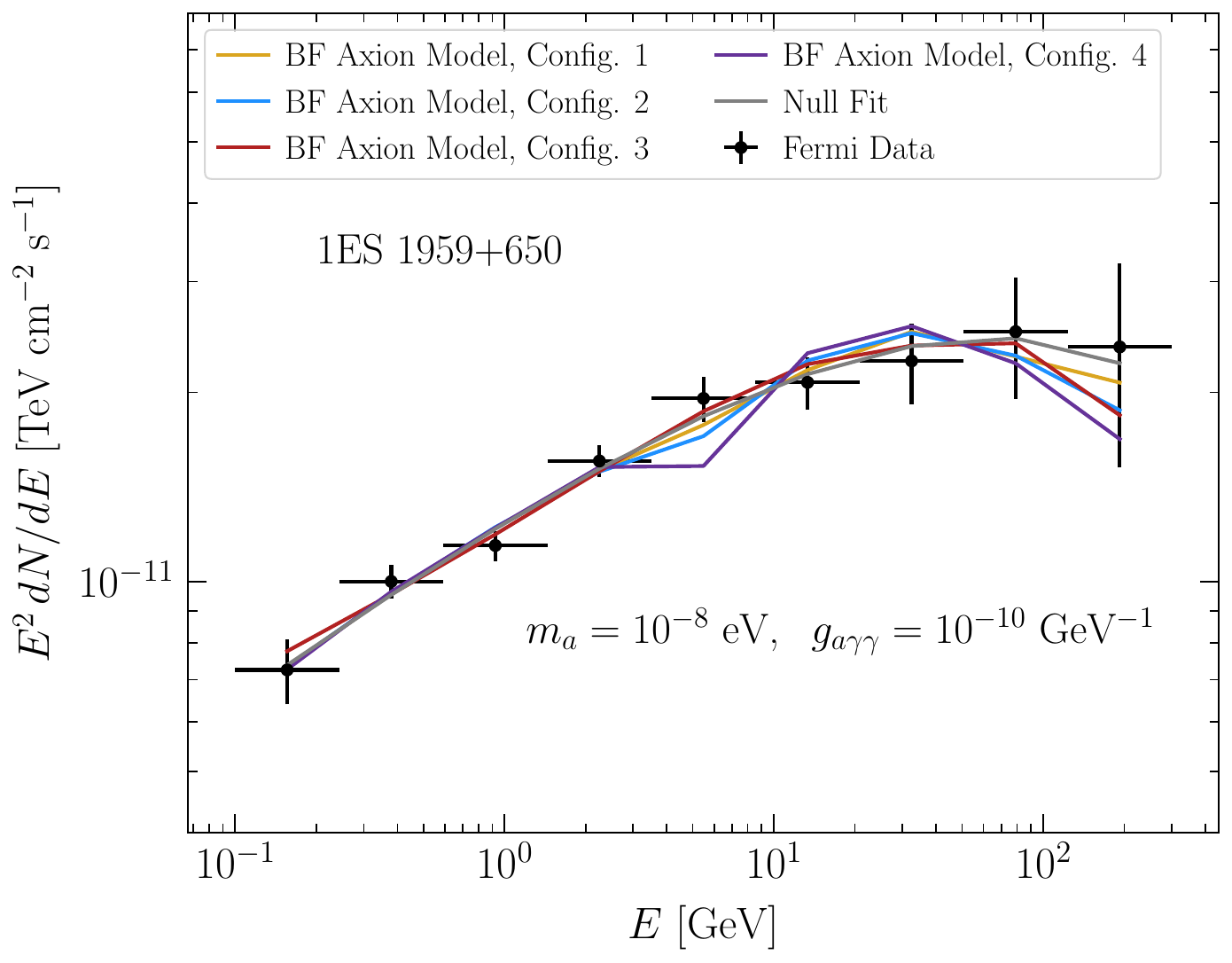}
\includegraphics[width=0.49\textwidth]{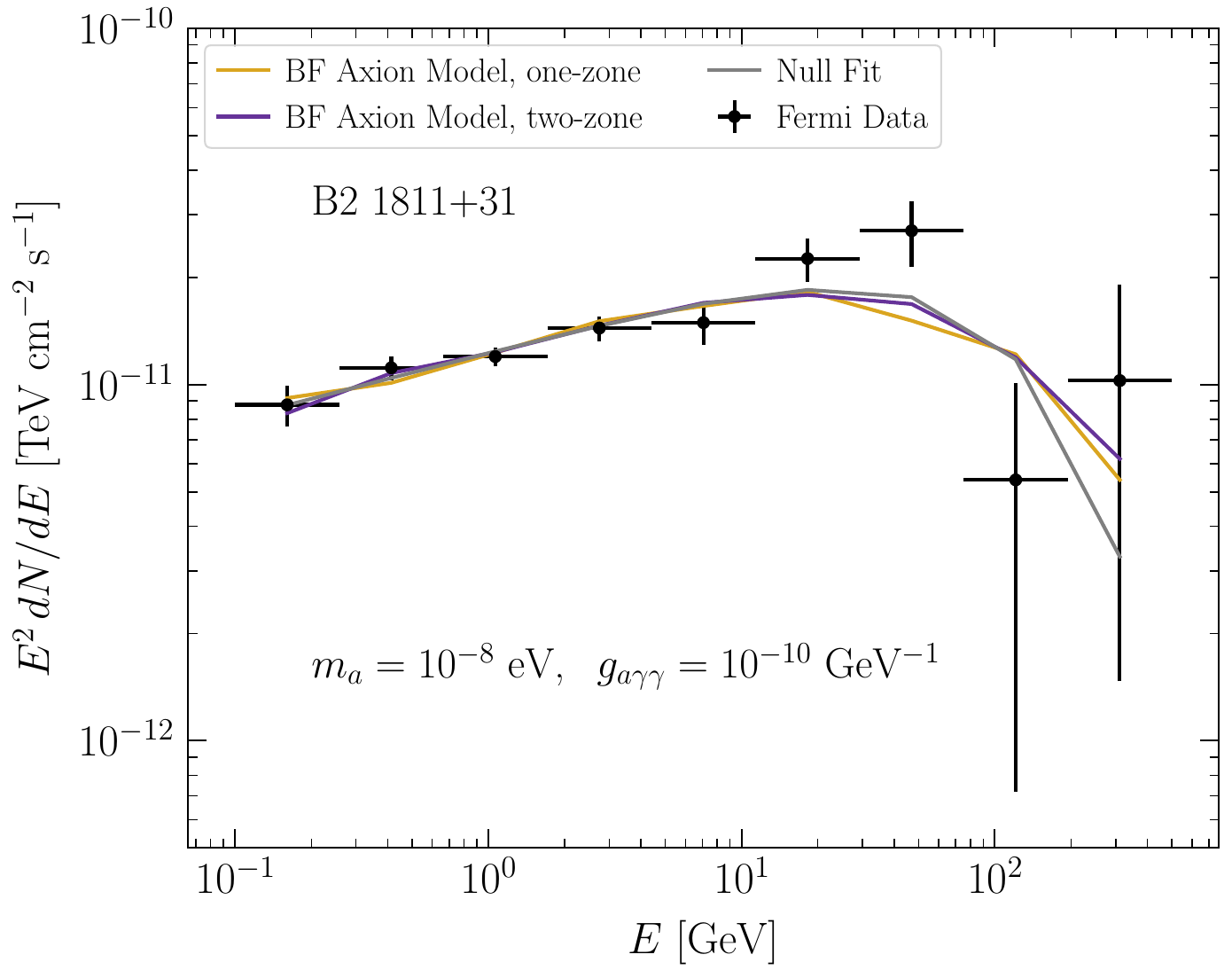}
\vspace{-0.4cm}
\caption{(Left) The Fermi-LAT spectral data for our blazar target 1ES 1959+650 compared to example axion-induced modulations over our four main fiducial configurations (see Table~\ref{tab:blazar_targets}) for the indicated mass and coupling. Also shown is the fit under the null hypothesis (gray). (Right) The same but for blazar target B2 1811+31 and its two fiducial models.}
\label{fig:spectra}
\end{figure*}

This convolved signal is then compared to the blazar data $\boldsymbol{d}$ in a spectral Gaussian likelihood 

\begin{equation}
    \mathcal{L}(\boldsymbol{d} | \{ m_a , \, \gagg, \, \boldsymbol{\theta}\}) = \prod_{i} \frac{1}{\sqrt{2 \pi} \sigma_i} \exp{\left[ -\frac{(d_i - \Phi_{\mathrm{sig}, i})^2}{2 \sigma_i^2} \right]}
\end{equation}
where $i$ refers to the energy bin, and $\sigma_i$ are the uncertainties on the observed flux from the data $\boldsymbol{d}$. Following frequentist standards, for any given $m_a$ and $\gagg$ we maximize the log-likelihood by profiling over our nuisance parameters, which in this case are our intrinsic modeling parameters $\boldsymbol{\theta}$. In Fig.~\ref{fig:spectra} we illustrate examples of the axion-induced modulation signal for our two main blazar targets, and compare to the spectral data of each.


To determine the significance of a potential axion discovery signal, one can construct the test statistic for the axion hypothesis over the null
\es{}{
    q(0) &\equiv -2 [ \ln{\mathcal{L}(\boldsymbol{d}|\{ m_a = 0, \, \gagg = 0\})} \\ 
    &- \ln{\mathcal{L}(\boldsymbol{d}|\{ \bar{m}_a, \, \bar{g}_{a\gamma\gamma}\})} ] \,,
}
where $\bar{m}_a$ and $\bar{g}_{a\gamma \gamma}$ are the global axion mass and coupling parameters which maximize the likelihood $\mathcal{L}$. Note that in all cases here and below, we also assume the nuisance parameter vector $\boldsymbol{\theta}$ which maximizes $\mathcal{L}$ for each signal parameter vector. In the absence of an axion discovery, we construct the TS suitable for setting constraints on the $(m_a, \, \gagg)$ axion parameter space. We define this as
\es{}{
    q(m_a ,\,\gagg) &\equiv -2 [ \ln{\mathcal{L}(\boldsymbol{d}|\{ m_a, \, g_{a\gamma\gamma}\})} \\ 
    &- \ln{\mathcal{L}(\boldsymbol{d}|\{ \bar{m}_a, \, \bar{g}_{a\gamma\gamma}\})}  ]
}
where again $\bar{m}_a$ and $\bar{g}_{a\gamma \gamma}$ are the global axion mass and coupling parameters which maximize the likelihood $\mathcal{L}$. A 95\% upper limit can be placed over our signal parameter combinations $(m_a, \, \gagg)$ given a 95\% threshold TS, $q_{95}$. 

Given that the spectral modulations induce a non-linear likelihood which is not smoothly varying in axion parameter space, one cannot use Wilks' theorem and other asymptotic formulae to infer a significance for $q(0)$ or to derive $q_{95}$~\cite{Cowan:2010js, Cowan:2011an}. Instead, we derive these quantities through Monte Carlo-simulated spectra for each of our blazars (and their systematic configurations), following the statistical prescription adopted from similar works, \textit{e.g.}, Refs.~\cite{Li:2020pcn, TheFermi-LAT:2016zue}. We accomplish this by drawing sample spectra from the observed quantities under the null hypothesis and observed uncertainties, and derive $q$ for each simulation, which we denote as $q_{\rm MC}$. Our ensembles of $q_{\rm MC}$ then form a non-Gaussian distribution from which we can gauge significances and extract the 95th percentile as our derived $q_{95}$, for each blazar, and for each particular systematic configuration. We remark that implicitly, as discussed in Ref.~\cite{TheFermi-LAT:2016zue}, there is the assumption that the signal distribution's derived $q_{95}$ is approximated by the null's, though Ref.~\cite{TheFermi-LAT:2016zue} verifies that the resulting limits from this procedure result in overcoverage of axion parameters, making them relatively more conservative.

\begin{figure*}[htb!]
\centering
\begin{minipage}{0.49\textwidth}
    \centering
    \includegraphics[width=\textwidth]{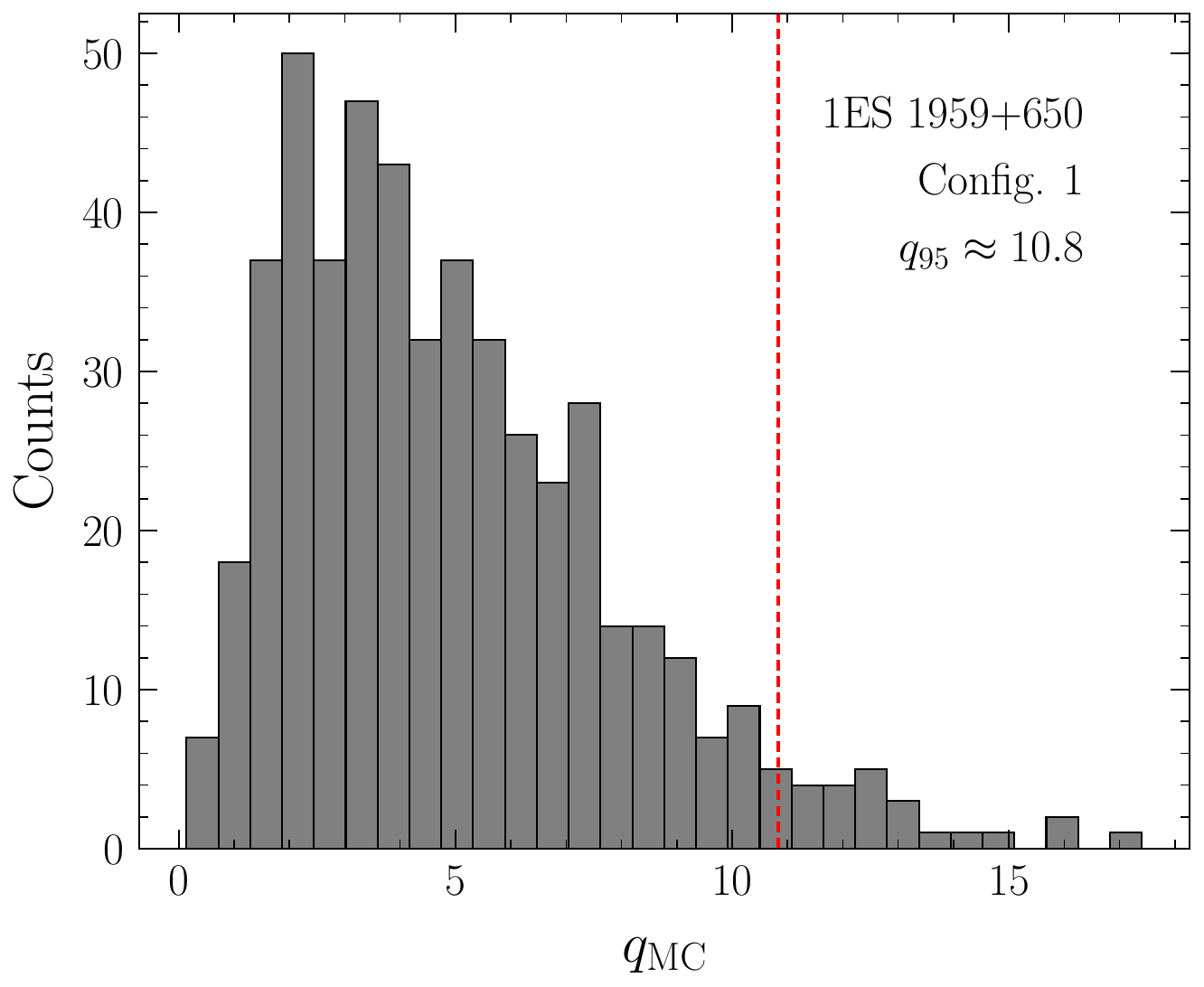}\\[-2pt]
\end{minipage}
\hfill
\begin{minipage}{0.49\textwidth}
    \centering
    \includegraphics[width=\textwidth]{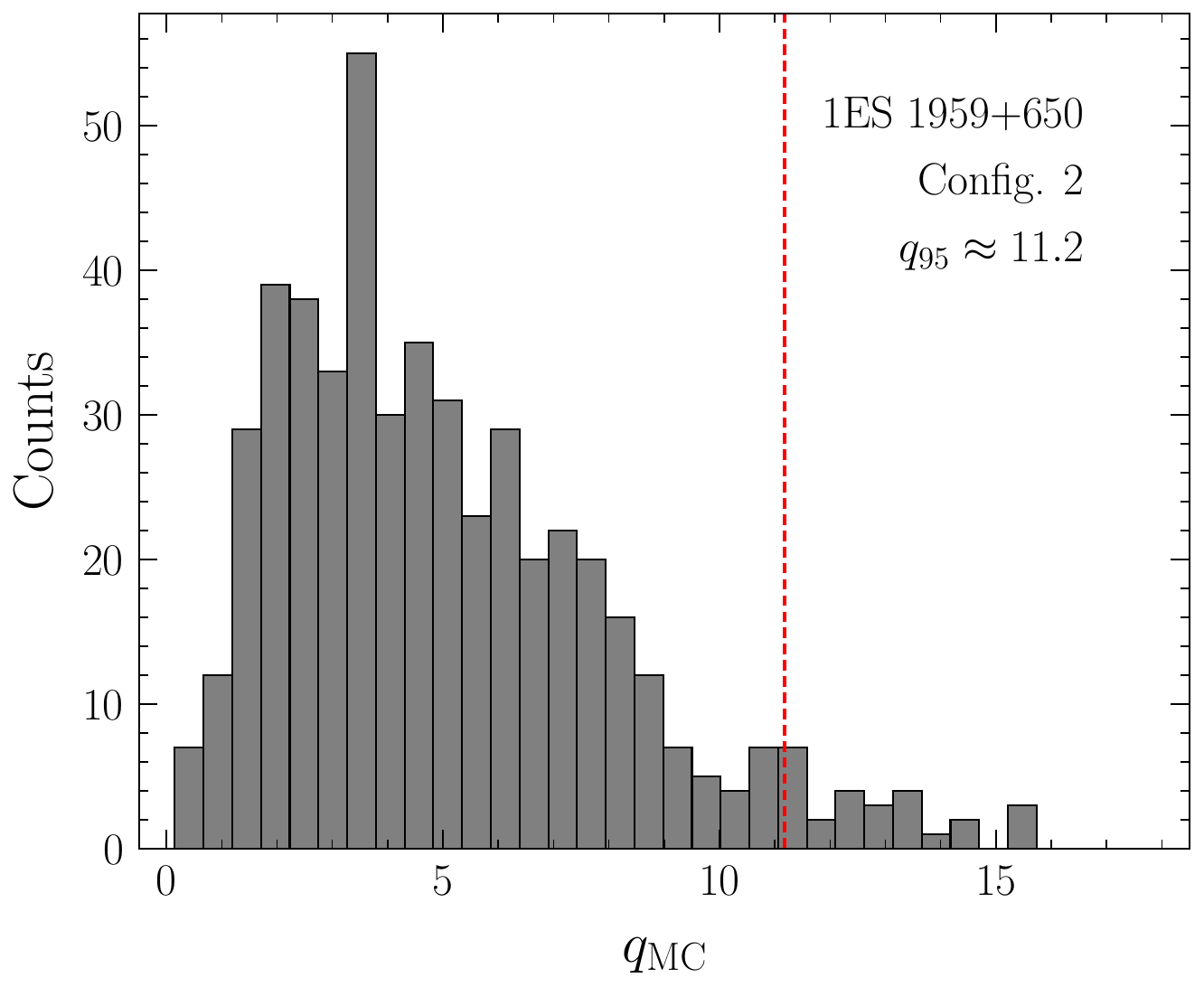}\\[-2pt]
\end{minipage}

\vspace{0.3cm}

\begin{minipage}{0.49\textwidth}
    \centering
    \includegraphics[width=\textwidth]{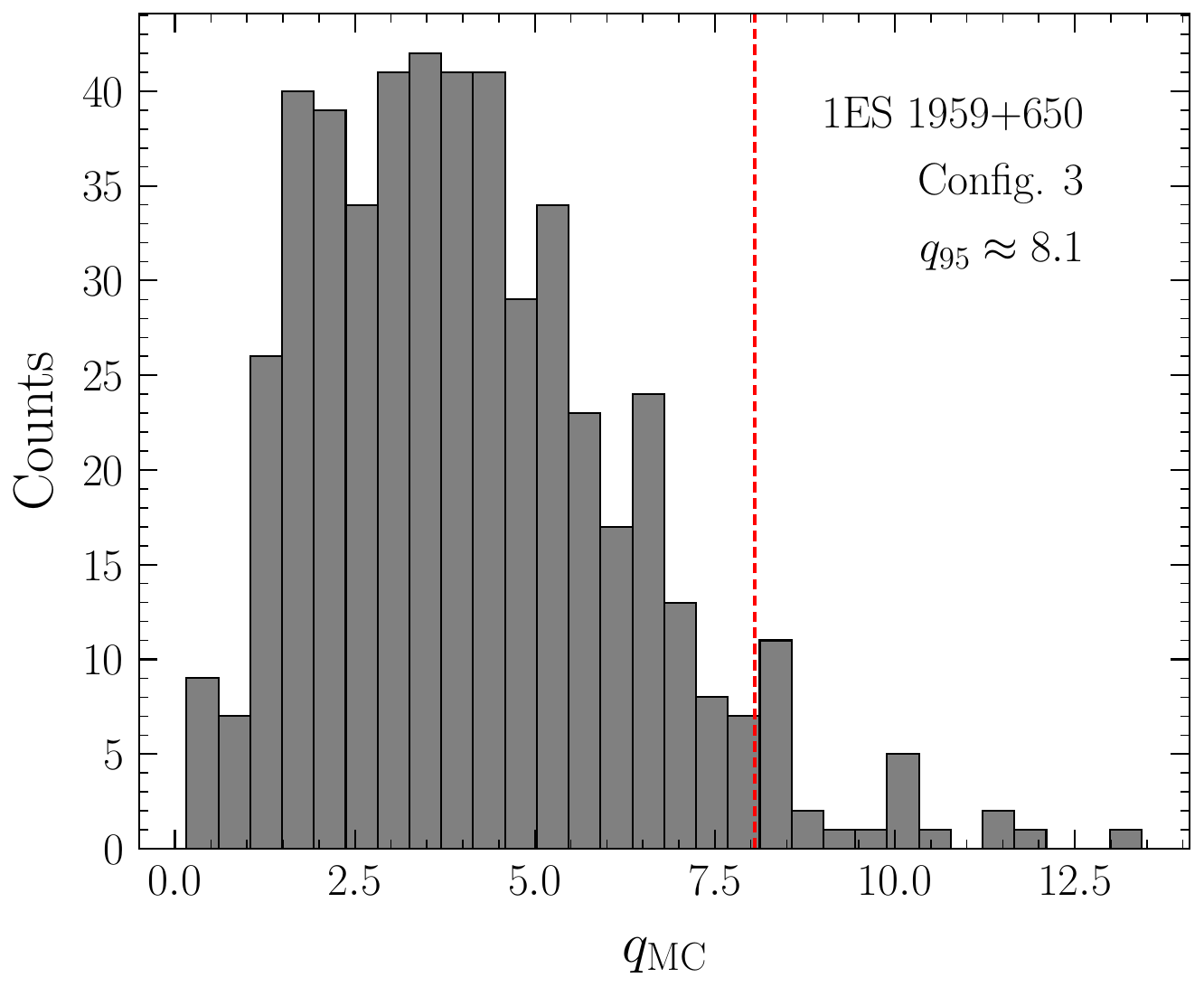}\\[-2pt]
\end{minipage}
\hfill
\begin{minipage}{0.49\textwidth}
    \centering
    \includegraphics[width=\textwidth]{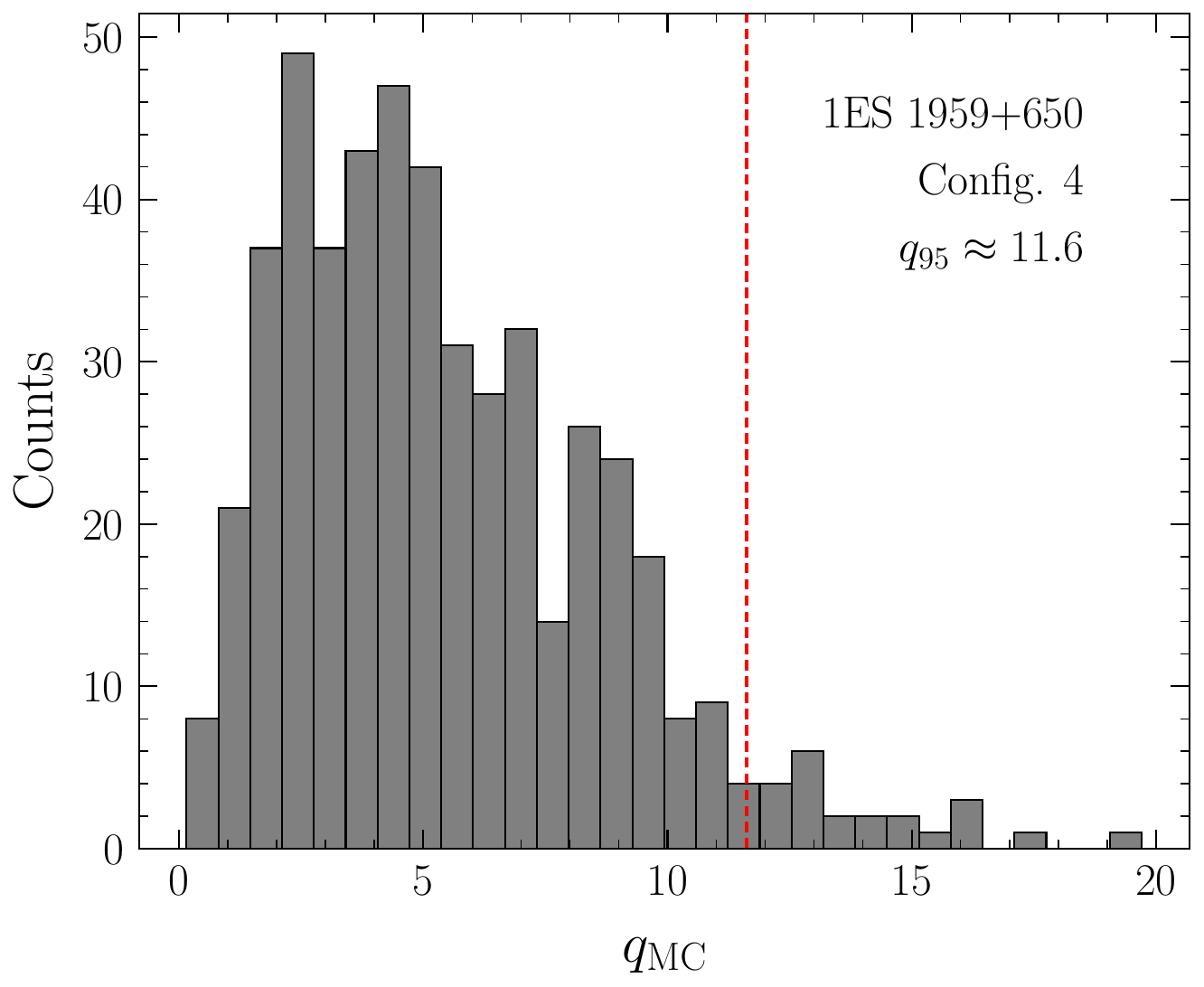}\\[-2pt]
\end{minipage}

\caption{The distribution of $q_{\rm MC}$ over our MC ensemble ($N = 500$) of systematic configurations of blazar target 1ES 1959+650 (see main text). We illustrate the derived $q_{95}$ value in dashed red, with its value indicated. The ordering top left, top right, bottom left, bottom right corresponds to configurations 1, 2, 3, 4, respectively.}
\label{fig:MC_1ES}
\end{figure*}

\begin{figure*}[!htb]
\centering
\includegraphics[width=0.49\textwidth]{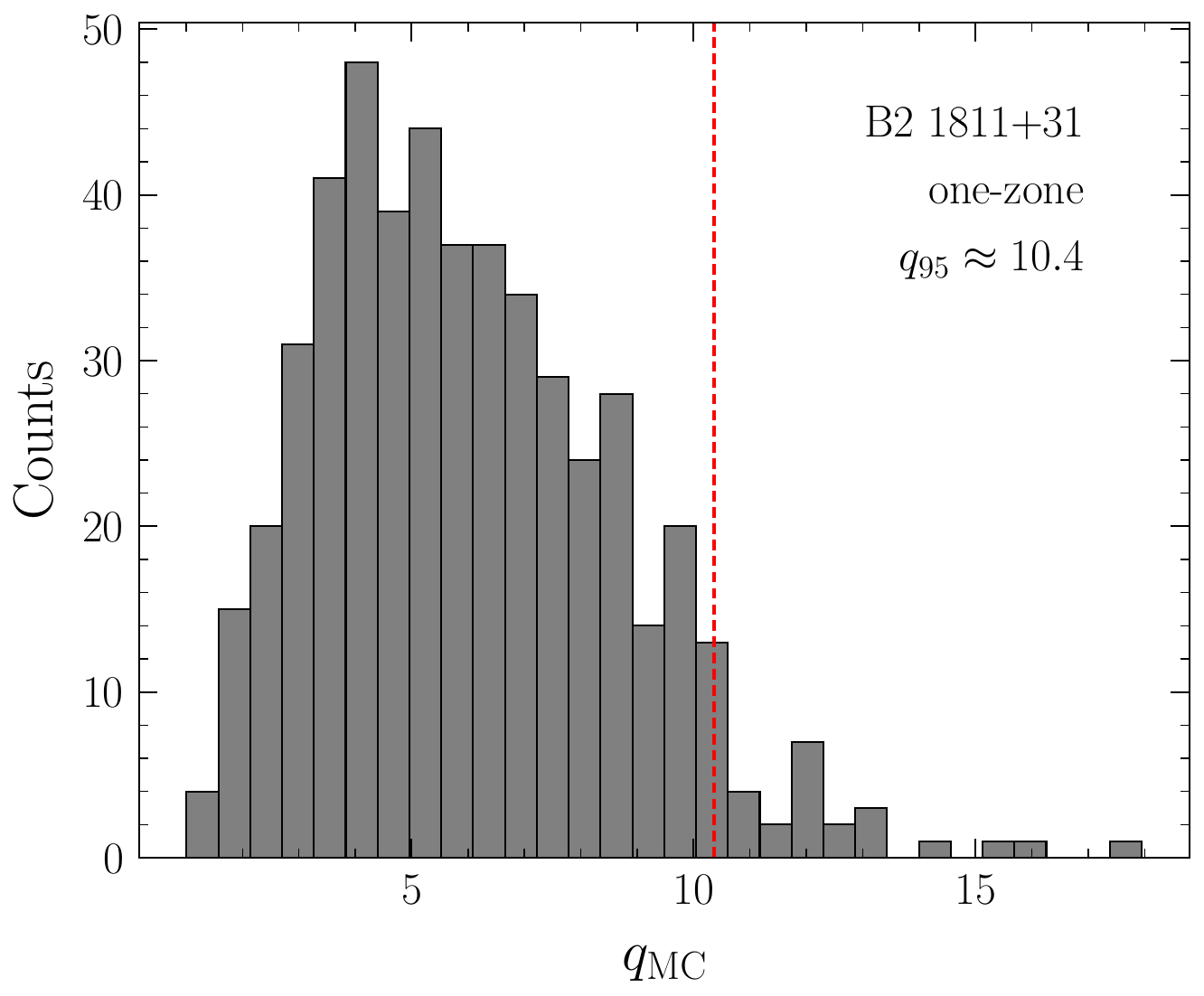}
\includegraphics[width=0.49\textwidth]{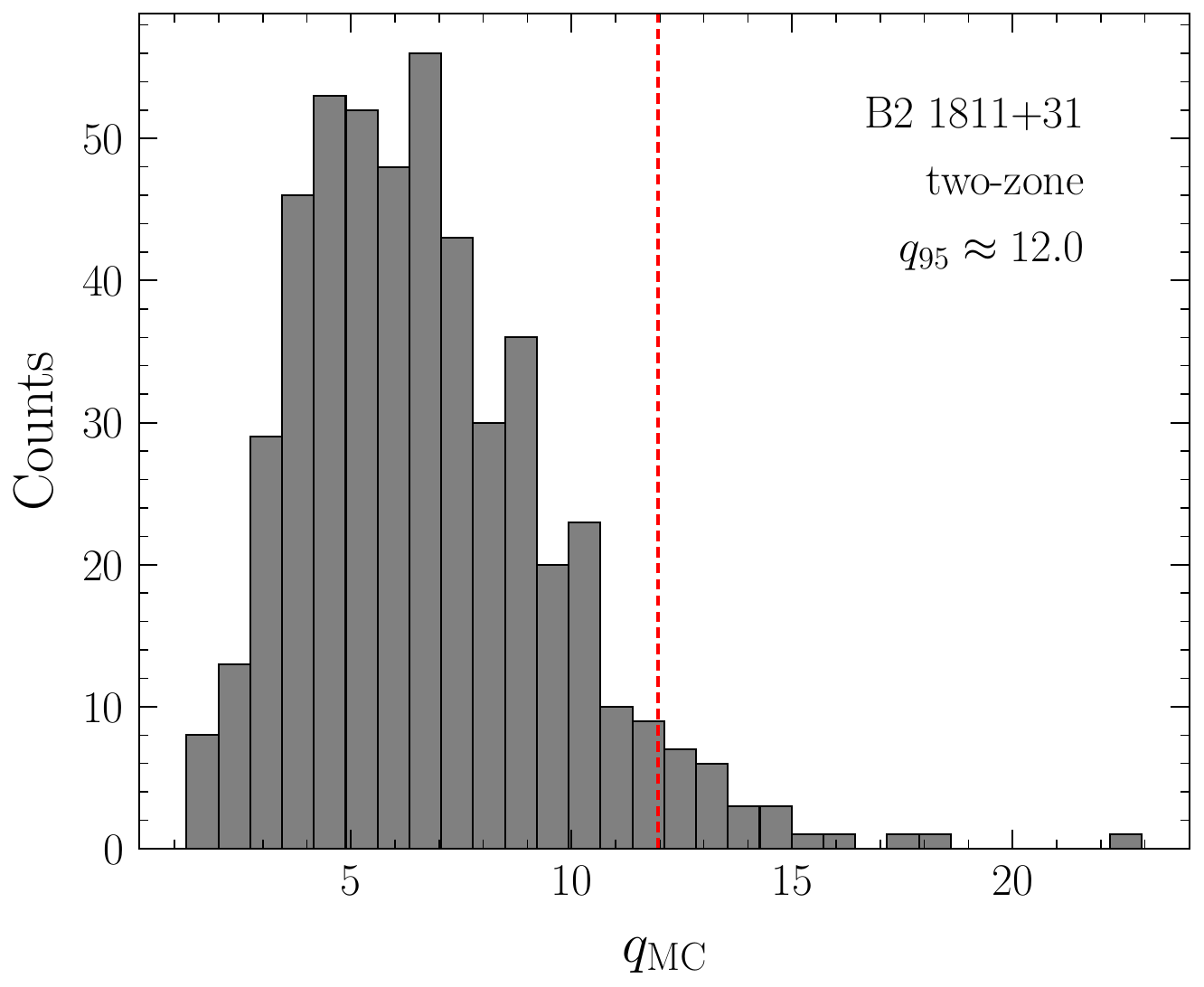}
\vspace{-0.4cm}
\caption{The same as Fig.~\ref{fig:MC_1ES} but for blazar target B2 1811+31. In the left panel we show the distribution of $q_{\rm MC}$ and the resulting $q_{95}$ value for the one-zone model without the core emission region, while in the right panel we include the core region in the two-zone model.}
\label{fig:MC_B2}
\end{figure*}

We illustrate the distribution of $q$ for the fiducial configurations of blazars 1ES 1959+650 and B2 1811+31 in Figs.~\ref{fig:MC_1ES} and~\ref{fig:MC_B2}, which are manifestly non-Gaussian. The thresholds $q_{95}$, which are indicated by the red dashed line, are markedly higher than the the equivalent that one would simply derive from Wilks' theorem.

\section{Results}
\label{sec:results_systematics}

\begin{figure*}[!htb]
\centering
\begin{minipage}{0.49\textwidth}
    \centering
    \includegraphics[width=\textwidth]{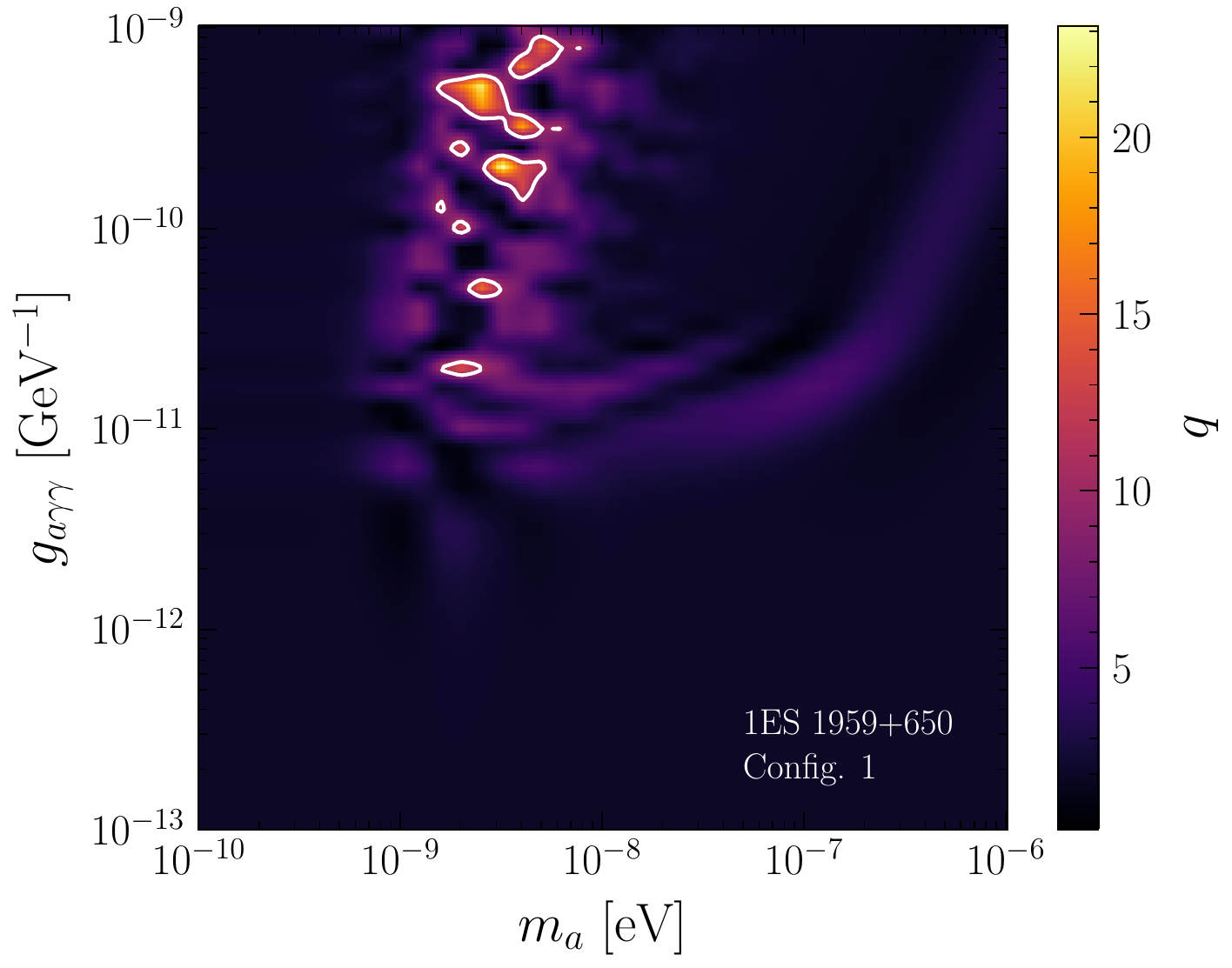}\\[-2pt]
\end{minipage}
\hfill
\begin{minipage}{0.49\textwidth}
    \centering
    \includegraphics[width=\textwidth]{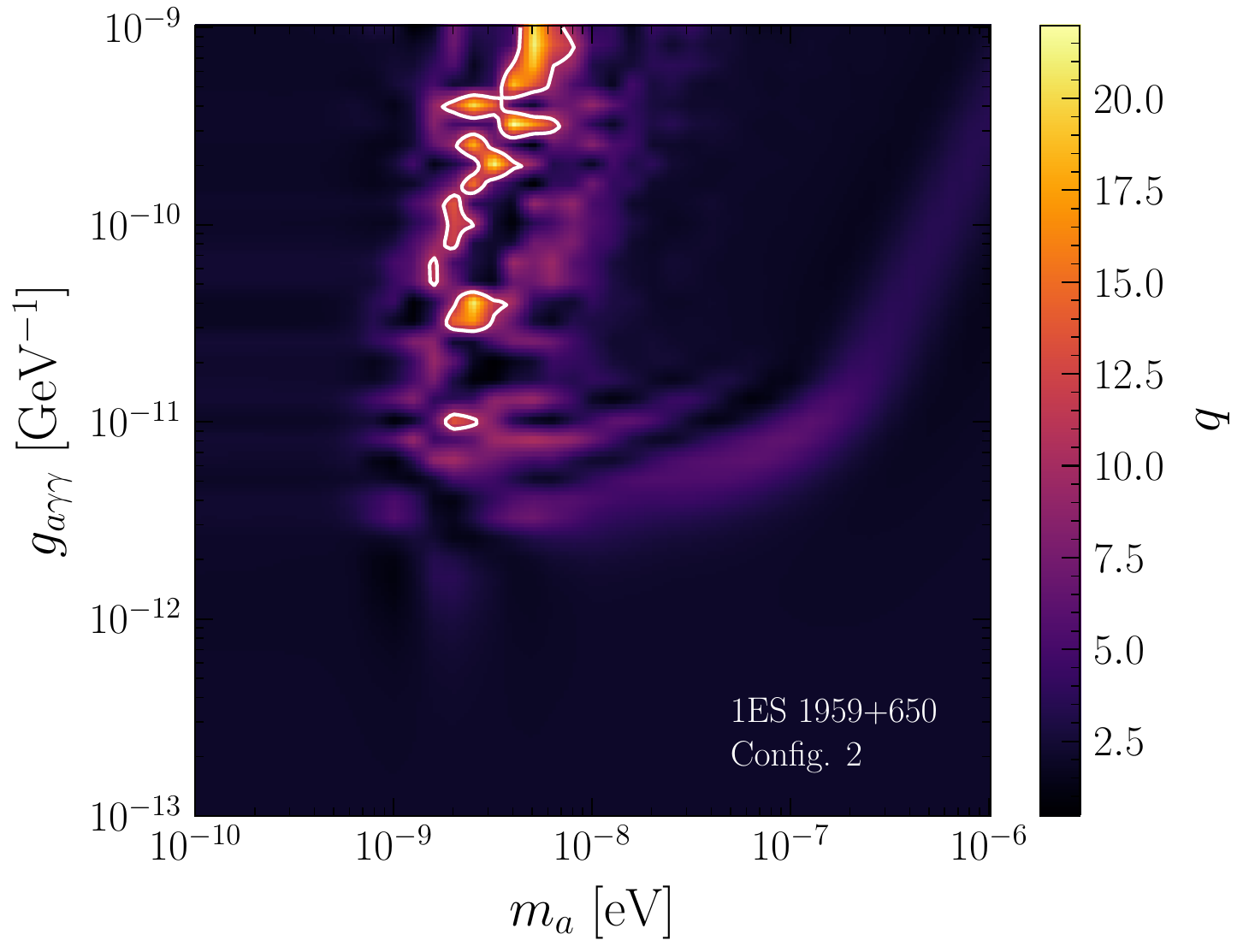}\\[-2pt]
\end{minipage}

\vspace{0.3cm}

\begin{minipage}{0.49\textwidth}
    \centering
    \includegraphics[width=\textwidth]{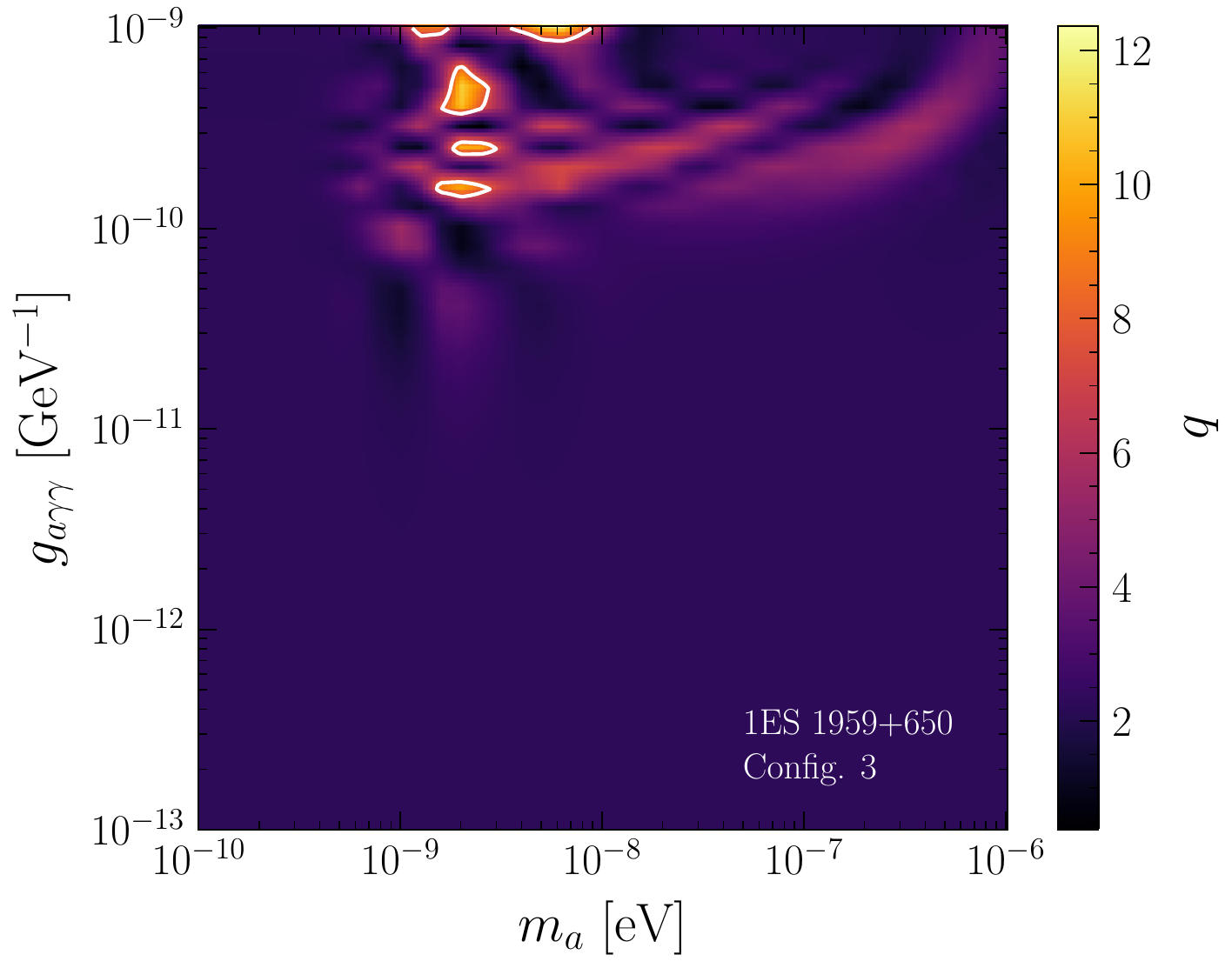}\\[-2pt]
\end{minipage}
\hfill
\begin{minipage}{0.49\textwidth}
    \centering
    \includegraphics[width=\textwidth]{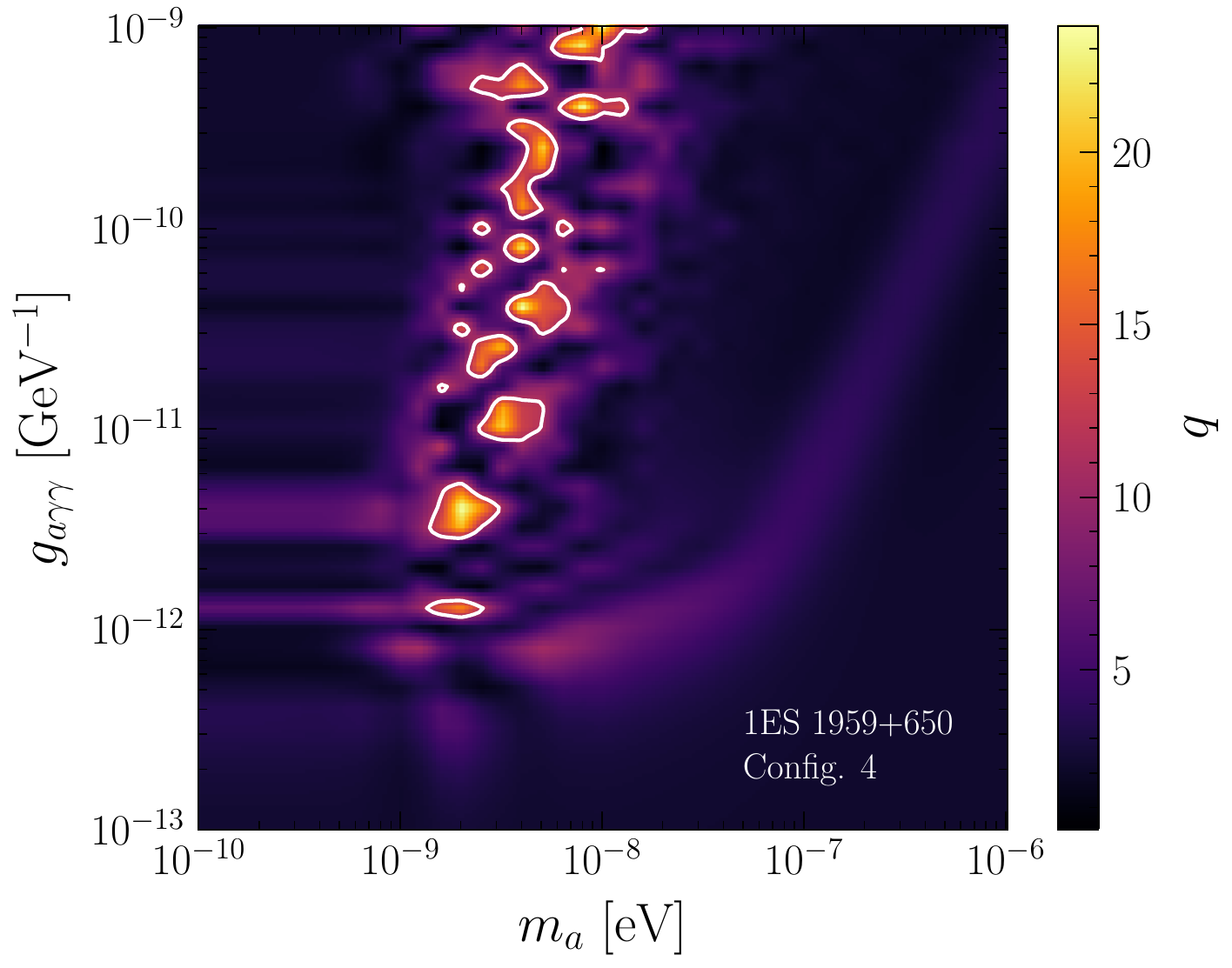}\\[-2pt]
\end{minipage}

\caption{The distribution of $q(m_a, \, \gagg)$ derived over our axion parameter space for our four astrophysical parameter configurations of blazar target 1ES 1959+650 (with the ordering of top left, top right, bottom left, bottom right referring to configurations 1, 2, 3, 4, respectively). We also illustrate, in white outline, the 95\% exclusion regions given our MC-derived $q_{95}$ value for each systematic configuration. }
\label{fig:qmap_1ES}
\end{figure*}

\begin{figure*}[!htb]
\centering
\includegraphics[width=0.49\textwidth]{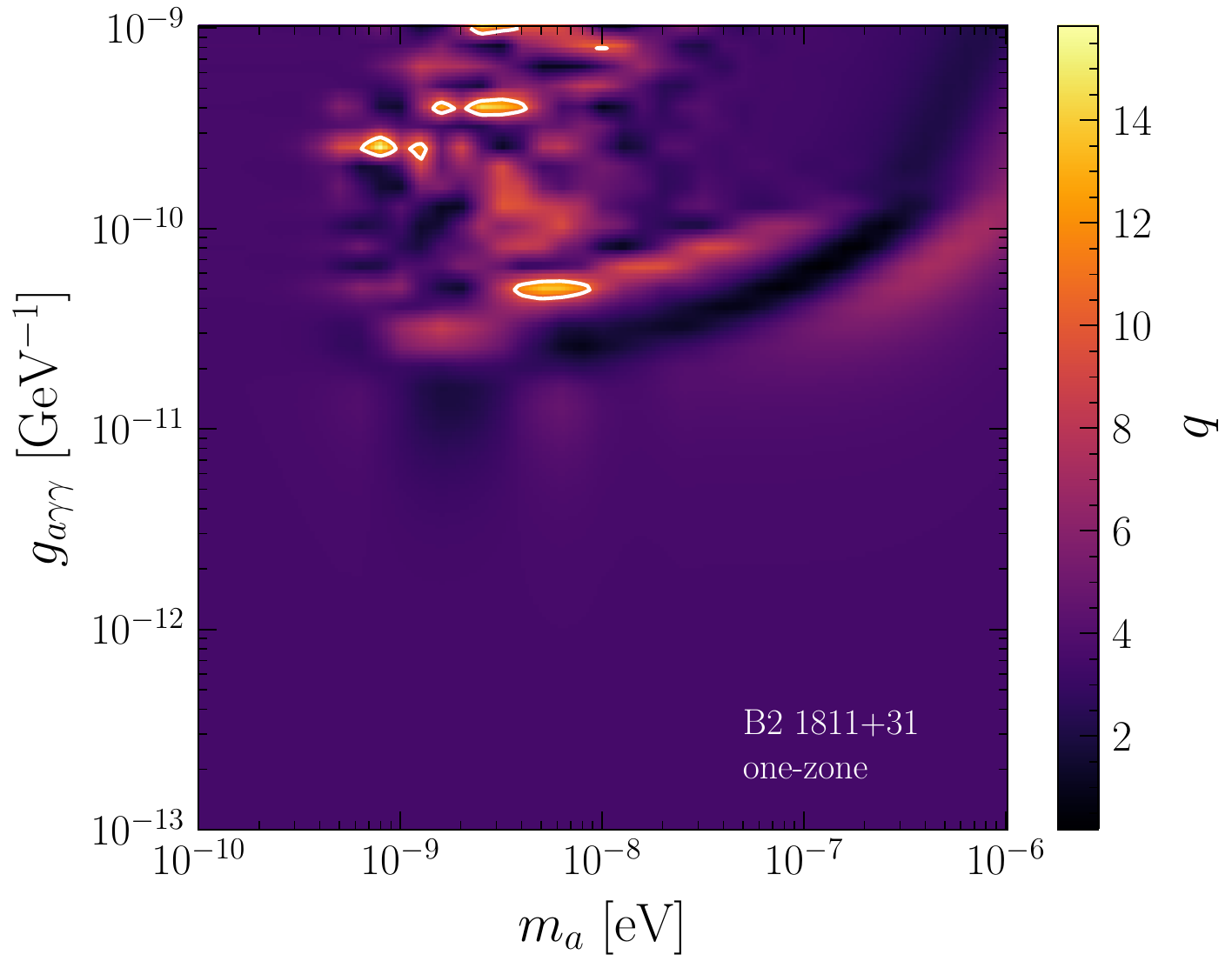}
\includegraphics[width=0.49\textwidth]{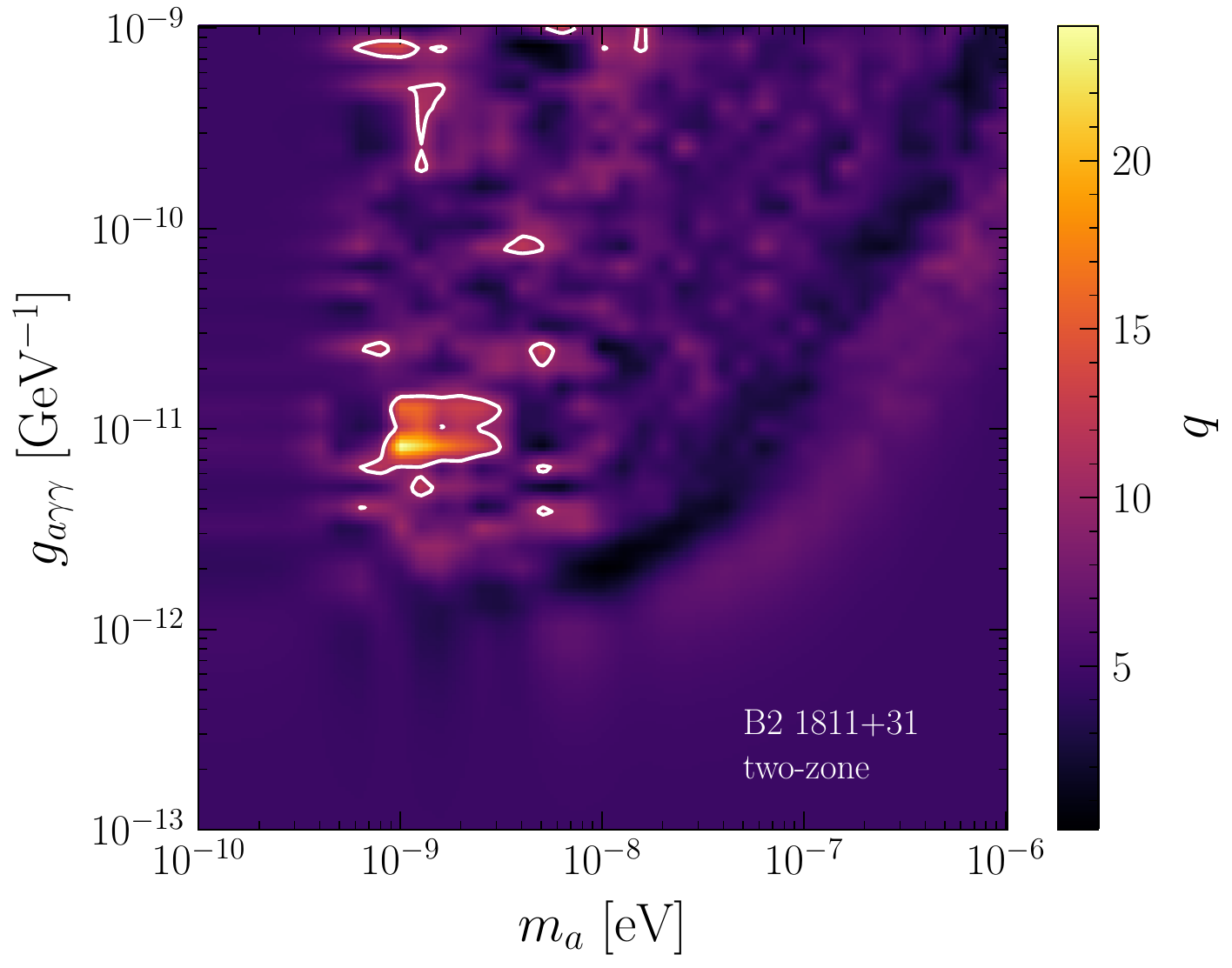}
\vspace{-0.4cm}
\caption{(Left) The distribution of $q(m_a, \, \gagg)$ derived over our axion parameter space for our one-zone configuration of blazar target B2 1811+31, omitting the core emission region. Our 95\% exclusion regions derived using $q_{95}$ are shown in white contours. (Right) The same but for the two-zone configuration, which includes the core region.}
\label{fig:qmap_B2}
\end{figure*}

We find no evidence for axions in any of the fiducial configurations for our two blazars considered in this work. Given $q_{95}$, we derive the configuration-dependent 95\% exclusion regions in $(m_a, \, \gagg)$ space for $q(m_a, \, \gagg) > q_{95}$. For each of our blazars and their corresponding configurations, we illustrate the distribution of $q(m_a, \, \gagg)$ over our axion parameter space together with the resulting 95\% exclusion regions in Figs.~\ref{fig:qmap_1ES} and~\ref{fig:qmap_B2}.  A representative subset of these exclusion regions is then compared to current constraints in the $m_a - \gagg$ plane in Fig.~\ref{fig:exclusions}. 

Among the configurations considered, the strongest sensitivity arises from 1ES 1959+650 under Config. 4, whose predicted modulations are strong enough to probe axions with masses between $10^{-9} - 10^{-8}$ eV down to couplings $\gagg \gtrsim 3 \times 10^{-12}$ GeV$^{-1}$. We emphasize again that, through our light curve analysis in App.~\ref{app:lc_analysis}, the parameters used in this configuration are indeed highly plausible. However, we also emphasize that the sensitivity for 1ES 1959+650 in general is highly dependent on the assumed configuration; Configs. 1 and 2 of 1ES 1959+650 are noticeably weaker, being sensitive to parameter space centered around $\gagg \sim 10^{-10}$ GeV$^{-1}$, and Config. 3 is the weakest of the main configurations, with no sensitivity below $\gagg \approx 10^{-10}$ GeV$^{-1}$, although we note again that the determination of Config. 3 parameters from MAGIC are also degenerate with hadronic models which would demand magnetic fields orders of magnitude higher. The spread of constraints across configurations illustrates the importance of astrophysical modeling uncertainties in interpreting these exclusion regions.

For blazar B2 1811+31, the two-zone configuration yields the most constraining results, where the presence of an additional core emission region enhances the overall modulation effects. With this configuration, one can probe parameter space volumes for $\gagg \gtrsim 5 \times 10^{-12}$ GeV$^{-1}$ for axion masses centered around $m_a \sim 10^{-9}$ eV. On the other hand, the one-zone model is poorly constraining, and although we again emphasize that this model is disfavored by multiwavelength observations and analyses (as discussed in Sec.~\ref{sec:blazar_env}), we analyze this one-zone model to show the impact of basic blazar modeling assumptions.

\section{Discussion}
\label{sec:discussion}

In this work, we conduct searches for axion-induced spectral modulations in the $\gamma$-ray band in high-energy blazar environments. We focus on blazars 1ES 1959+650 and B2 1811+31, using Fermi-LAT data of flare states of each respective target, and calculate the modulations expected under the axion hypothesis in each unique blazar environment. Comparing to the data, we find no evidence for axions, and instead set configuration-dependent 95\% exclusion regions on the axion mass and coupling $(m_a,\gagg)$ parameter space. 

The sensitivity of these constraints depends significantly on the assumed properties of the blazar environment. In the most optimistic blazar model for 1ES 1959+650, the analysis is sensitive to couplings as low as $\gagg \gtrsim 3 \times 10^{-12}$ GeV$^{-1}$ for axion masses $m_a \sim 10^{-9}$ eV, though other plausible configurations more conservatively probe regions closer to $\gagg \sim 10^{-10}$ GeV$^{-1}$ for axion masses between $10^{-9}$ eV $\lesssim m_a \lesssim 10^{-8}$ eV. The spread in constraints among configurations reflects the sensitivity of spectral modulation searches in general to blazar environment assumptions, and so it is difficult to claim robust exclusions based on one configuration alone; in the inset of Fig.~\ref{fig:exclusions}, we highlight more restrictive regions of parameter space simultaneously excluded by multiple configurations for each blazar, providing a more realistic indication of the parameter space excluded in this analysis when systematic modeling uncertainties are fully considered.


We explore these systematics through our modeling, which mostly attempt to place a measure of uncertainty on the two parameters most relevant for these spectral modulations, which are the blazar magnetic field $B$ and emission region size $R_b$, both of which strongly affect the spectral distortions associated with axions in the blazar environment. Future observations of these blazars would provide valuable data on these key parameters which would help to more accurately probe the axion phenomenology explored in this work. Specifically, future progress will require a more accurate mapping between the observed variability and the underlying magnetic-field coherence structure relevant for axion-photon mixing. Additional environments and other blazar targets could also be potentially interesting; for example, we also explored blazar target PKS 0625-354 whose $\sim$3 G magnetic field appeared promising for axion modulation searches, though its blob size, which is a couple orders of magnitude smaller than our main targets in this work, significantly limit the ability of this target to constrain axion parameters in the ranges of interest (\textit{i.e.} $\gagg \lesssim 10^{-9}$ GeV$^{-1}$). In addition, photon statistics can affect the sensitivity (see App. \ref{app:phase_sel}), and future searches should preferentially focus on bright sources with extended observational periods. 

We also speculate that other physical scenarios, such as particle emission models beyond the leptonic and hadronic ones mentioned here, would also have potentially interesting applications toward further probing axions or even other related particle phenomena such as high-energy neutrino production. More broadly, we finally note that population-level analyses of various blazars in a joint analysis may offer a complementary path forward. We leave such studies to future work.

\begin{acknowledgements}
{
{\it
We thank Matteo Cerruti, Yujin Park, Nick Rodd, and Ben Safdi for helpful discussions and conversations. We especially thank Nick Rodd for particularly enlightening suggestions and comments. The work of O.N. is supported in part by the DOE award DESC0025293, as well as the NSF Graduate Research Fellowship Program under Grant DGE2146752. This research used resources of the National Energy Research Scientific Computing Center (NERSC), a U.S. Department of Energy Office of Science User Facility located at Lawrence Berkeley National Laboratory, operated under Contract No. DE-AC02-05CH11231 using NERSC award HEP-ERCAP0023978. The work of A.G.D.M. is supported in part by the Italian INFN program on Theoretical Astroparticle Physics. A.G.D.M. is grateful to Lawrence Berkeley National Laboratory for their kind hospitality during the initial phases of this work.
}
}
\end{acknowledgements}

\appendix

\section{Light Curve Analysis}\label{app:lc_analysis}

In this section we give details regarding our own determination of the blob size $R_b$ of blazar 1ES 1959+650 by performing a light curve analysis on our data. Adopting the one-zone SSC model, the $\gamma$-ray is assumed to be produced by a spherical plasma blob of radius $R_b$. The blob contains a spatially homogeneous, isotropic population of electrons and a uniform magnetic field, with size $R_b$ constrained by variability through 

\begin{equation}
    R_b\lesssim \frac{c \, t_{var} \, \delta}{1+z} \,, \label{blobcausality}
\end{equation}
with a known redshift $z$. Taking a typical Doppler factor for blazars $\delta \sim 15-30$~\cite{2010IJMPD..19..841T}, this expression produces an upper limit on $R_b$ once the variability time scale $t_{var}$ is measured.

\begin{figure*}[!htb]
    \centering
    \includegraphics[width=\textwidth]
    {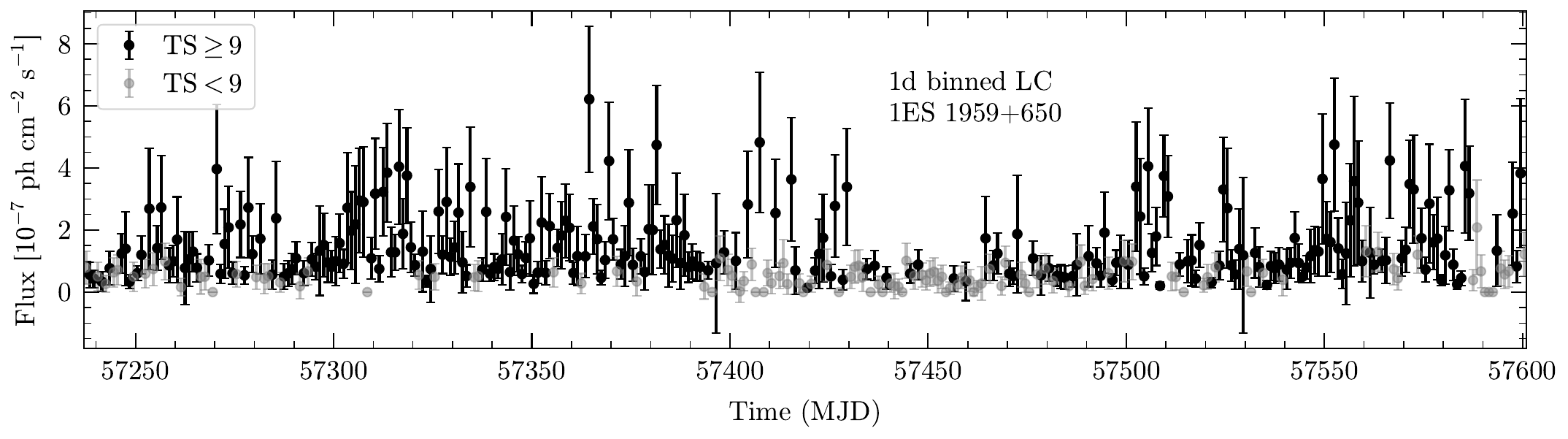}
    \includegraphics[width=\textwidth]
    {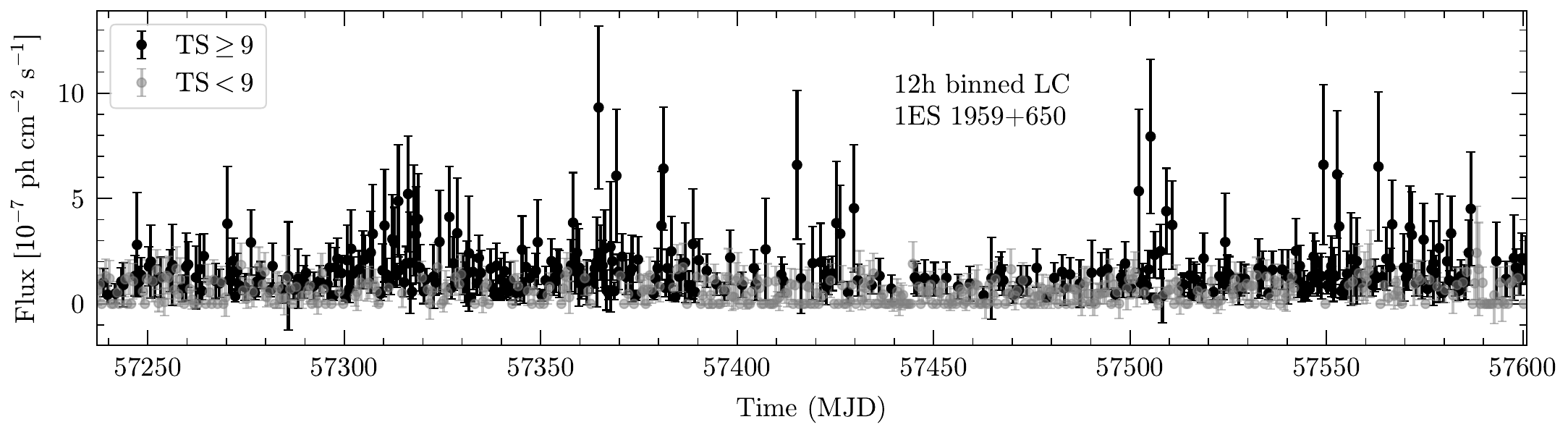}
    \caption{Fermi-LAT Pass 8 light curves of 1ES 1959+650 (MJD 57237-57601). (Top) The 1-day binned flux. Black points denote time bins with TS $\geq$ 9, retained as reliable flux measurements, while gray points correspond to bins with TS $<$ 9. Vertical bars represent 1$\sigma$ uncertainties. The variability analysis uses only the TS $\geq$ 9 points. (Bottom) The same but for the 12-hour binned flux. We note a higher fraction of less robust TS $<$ 9 measurements with shorter time bins.}
    \label{fig:lc_analysis}
\end{figure*}

We determine $t_{var}$ by analyzing the integrated light curves. In the case of 1ES 1959+650, we generate the light curve $L(t)$ from Fermi-LAT Pass 8 data using standard event selections and binned likelihoods. The photon flux is extracted in uniform space time bins, and only bins with TS $\geq9$ are retained as reliable measurements, consistent with previous standard light curve analyses~\cite{2023ApJS..265...31A}. We then quantify $t_{var}$ by identifying the shortest time between adjacent bins where significant flux changes occur: 

\begin{equation}
    t_{var}=\min \{\Delta t| |L(t+\Delta t)-L(t)|>N\sigma\} \,,
\end{equation}
where $\sigma$ is the observed flux uncertainty and $N$ is typically taken as 2-3. In our analysis, the 1-day light curve shows the most significant flux change at $N \approx 2.68$. The 12-h light curve shows a marginal jump of $N \approx 2$, while the 6-h and 3-h show no jumps with $N \gtrsim 2$ at all. We therefore take $t_{var}\simeq1$ day as the characteristic variability timescale for the flare phase considered. We show examples of our light curves in Fig.~\ref{fig:lc_analysis} for 1d and 12h bins, noting also that at shorter time bins the fraction of measurements with TS $<$ 9 noticeably increases.

Substituting this value into Eq.~\ref{blobcausality}, together with a typical doppler factor $\delta\sim 15-30$, yields an upper limit on the blob size of order 

\begin{equation}
    R_b\lesssim 3\times 10^{16} \, {\rm cm} \,,
\end{equation}
which we find is consistent within an $\mathcal{O}(1)$ factor of the $R_b$ found in Ref.~\cite{Patel:2017mrv}, justifying our adoption of that $R_b$ for our Config. 4. Note that this constraint does not restrict how small the emission region can be. Possible ways of obtaining lower limits on $R_b$ have been discussed in other works such as Ref.~\cite{Tavecchio:1998xw}, including arguments based on requiring moderate Doppler factors combined with  transparency conditions. However, these are highly model-dependent, which we do not consider robust enough to be implemented in our analysis.

Assuming that the jet has a conical geometry with an opening angle $\theta$ (\textit{e.g.} from Ref.~\cite{2013EPJWC..6106008S}), the radius of the blob is geometrically related to its radial position $r_0$ from the center of the black hole

\begin{equation}
    R_b \sim r_0\theta \,,
\end{equation}
such that the variability-derived upper limit on $R_b$ can also be translated into an upper limit on $r_0\lesssim 5\times 10^{17}$cm.

\section{Phase Selection}
\label{app:phase_sel}

\begin{figure*}[!bht]
\centering
\includegraphics[width=0.49\textwidth]{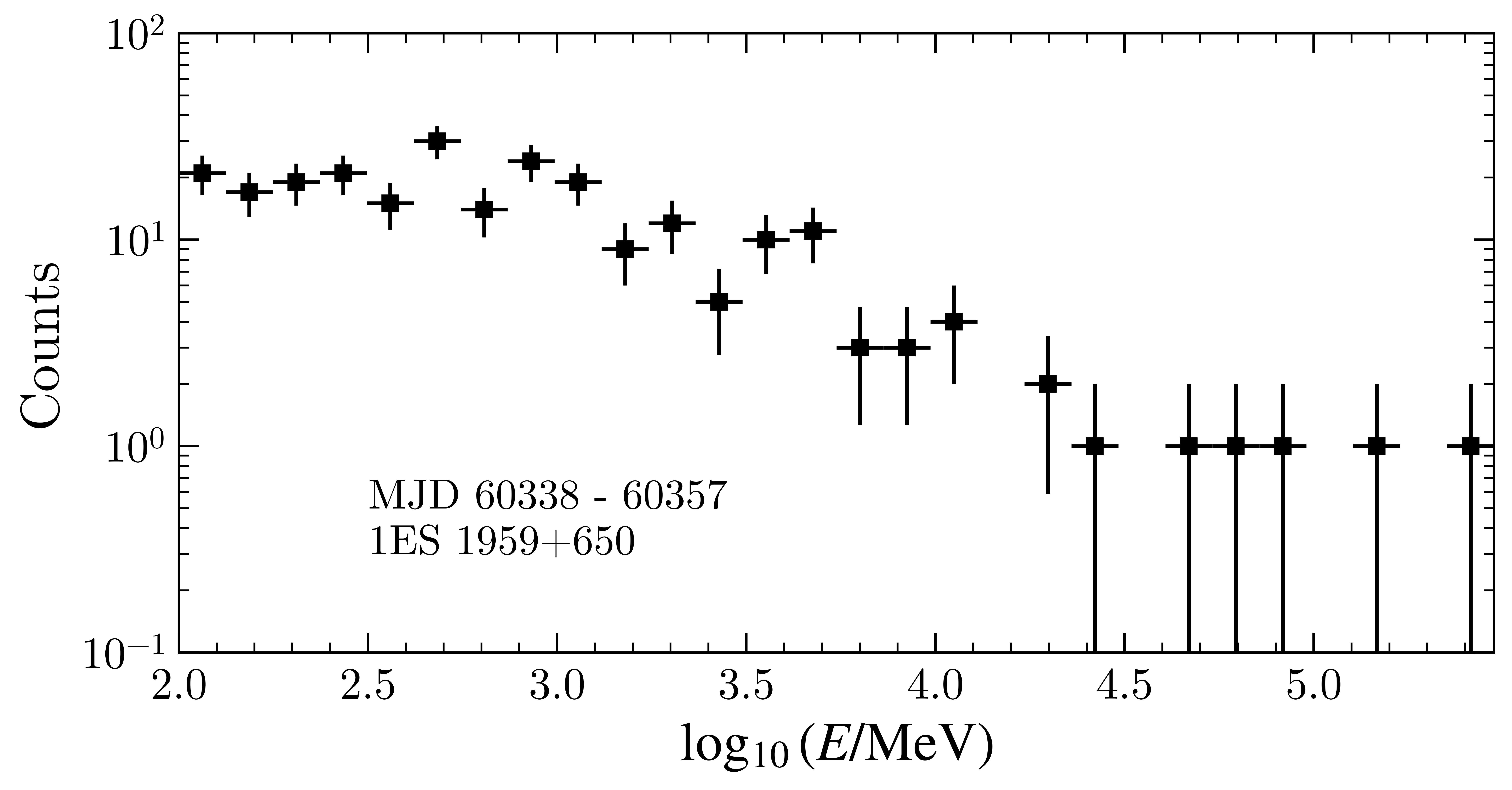}
\includegraphics[width=0.49\textwidth]{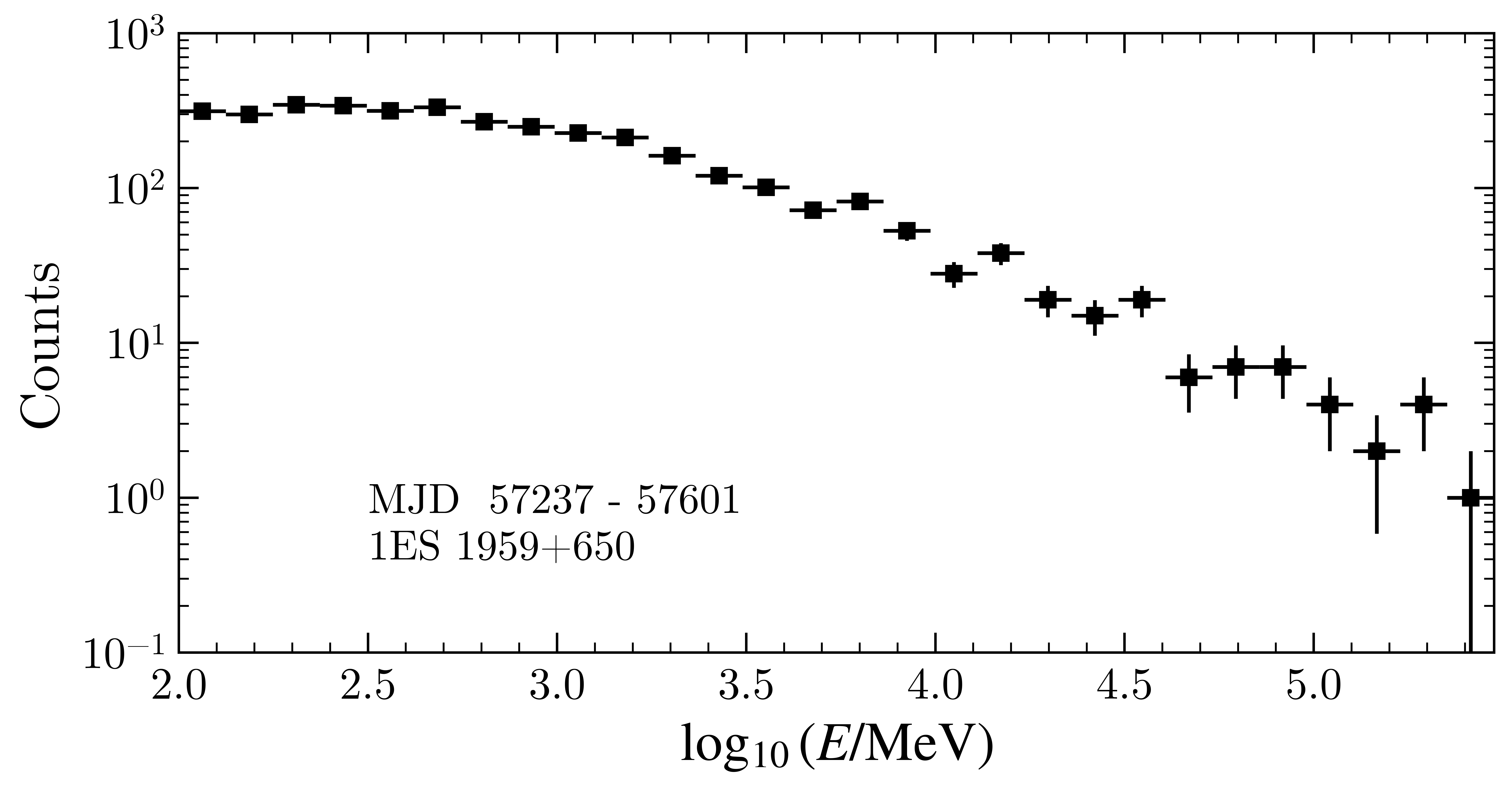}
\vspace{-0.4cm}
\caption{Illustrations of the initial photon counts spectra for 1ES 1959+650 obtained from the \texttt{Fermipy} binned likelihood analysis. (Left) Counts for MJD 60338 - 60357 (one month). (Right) Counts for MJD  57237 - 57601 (one year). The energy binning is fixed by the configuration file with default \texttt{binsperdec=8}. Error bars represent Poisson uncertainties.}
\label{fig:photoncount}
\end{figure*}

The $\gamma$-ray spectra in this study are obtained by integrating events over finite time intervals, the choices of which are non-unique for variable sources like blazars. Since photon-axion mixing is affected by local physics at the source, time-window selection must suppress variability-induced artifacts without imposing unjustified assumptions about emission physics.

For instance, if the phase window is too short, photon counts become insufficient. In this regime, photon fluctuations dominate individual energy bins, and bin-to-bin fluctuations could generate spectral features even when the true intrinsic spectrum is smooth, thereby reducing sensitivity. On the other hand, enlarging the time window improves photon statistics and should not wash away a genuine axion signal. The primary limitation of long time windows is that the intrinsic spectrum may evolve so drastically that a single set of blazar parameters can no longer describe its physical state at emission adequately.

However, the notion of a distinct blazar state is not well-defined. The parameters of one-zone emission models can be degenerate and not directly observable, and modest changes in flux or spectral shape do not necessarily imply different physical configurations. This limitation is widely acknowledged in blazar modeling and variability studies. We therefore do not attempt to identify distinct physical states. Instead, we adopt the approximate but reasonable criterion that if the flux level remains sufficiently stable, a time-averaged spectrum can be meaningfully used to derive constraints on axion signals.


As a cross-check of the time-window dependence, we examine a one month dataset of 1ES 1959+650 from 2024 (MJD 60338 - 60357) using the same methodology pipeline as for the 2015-2016 flare. The corresponding photon count spectra for both phases are shown in Fig.~\ref{fig:photoncount}, illustrating that the shorter time interval is relatively more photon-limited, with more visible bin-to-bin Poisson fluctuations, which we choose not to analyze in this work. More exactly, we emphasize that in our fiducial analysis we strive for time windows which minimize the possibility of Poisson fluctuations being comparable to or larger than the characteristic oscillation amplitudes of the photon survival probability, \textit{e.g.} Fig.~\ref{fig:prob_survival}.

\section{Spectral Modeling Variations}
\label{app:spectra_modeling}

\begin{figure*}[!htb]
\centering
\includegraphics[width=0.49\textwidth]{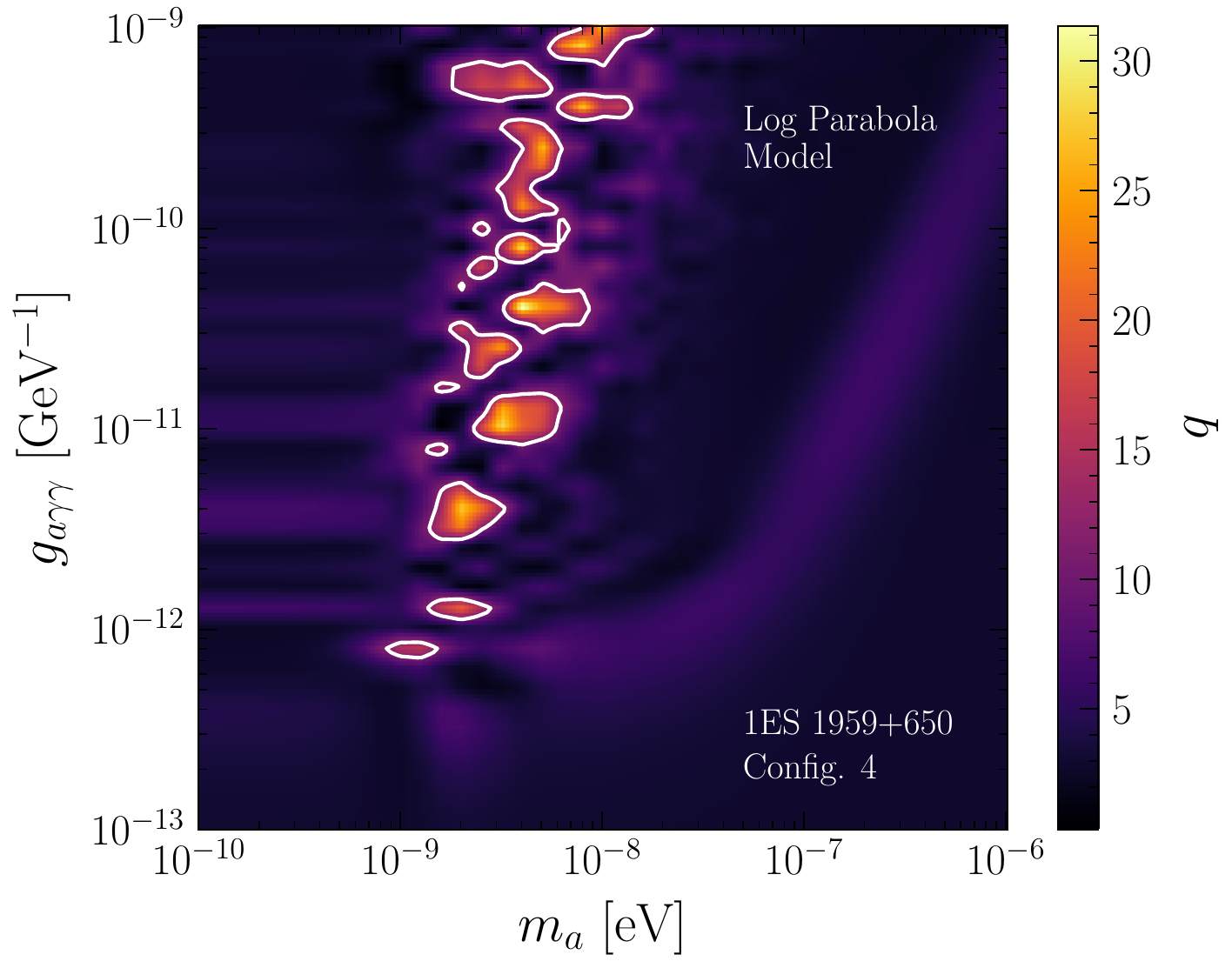}
\includegraphics[width=0.49\textwidth]{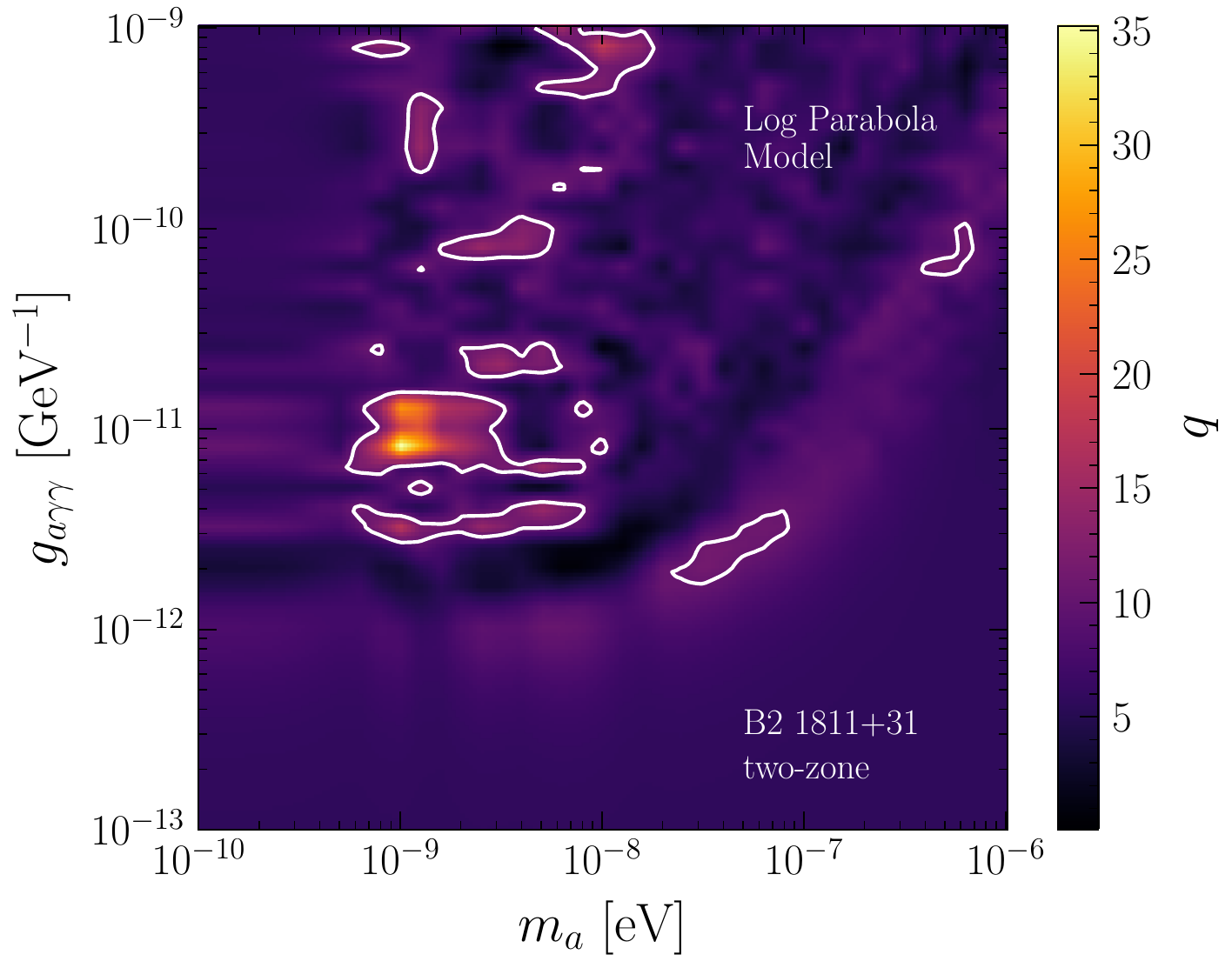}
\vspace{-0.4cm}
\caption{(Left) The distribution of $q(m_a, \, \gagg)$ and the 95\% constraints derived over our axion parameter space for Config. 4 of blazar target 1ES 1959+650, assuming the log-parabola intrinsic spectral model. (Right) The same but for the two-zone configuration of blazar B2 1811+31.}
\label{fig:logp}
\end{figure*}

In the main text we describe the intrinsic emission spectra from the canonical SEPL model, with four nuisance parameters $(\phi_0, \alpha, \beta, E_{\rm cut})$. As another test of the systematic uncertainties underlying our analyses, we perform the same analyses for blazars 1ES 1959+650 and B2 1811+31 assuming another intrinsic spectral model, the log-parabola model (considered in, \textit{e.g.}, Ref.~\cite{Li:2021gxs}). It is parameterized as follows:

\begin{equation}
\begin{alignedat}{1}
\phi(E) &= \phi_0 \left(\frac{E}{E_0}\right)^{-\alpha-\beta \log(E/E_0)}
\end{alignedat}
\end{equation}
with nuisance parameters $(\phi_0, E_0, \alpha, \beta)$. For the sake of demonstration, we choose to examine Config. 4 for 1ES 1959+650 and the two-zone model for B2 1811+31 (although we have checked other configurations lead to similar conclusions). As illustrated in Fig.~\ref{fig:logp}, for both our blazars we find similar 95\% constraints on the axion parameter combination $(m_a, \gagg)$, with marginally stronger constraints for B2 1811+31 under the log-parabola model compared to the SEPL model in the main text. We conclude here that these two conventional descriptions of the underlying spectra of both blazars imply similar axion constraints.

\section{Signal Injection Test}
\label{app:sig_inj}
\begin{figure}[!htb]
    \centering
    \includegraphics[width=0.49\textwidth]{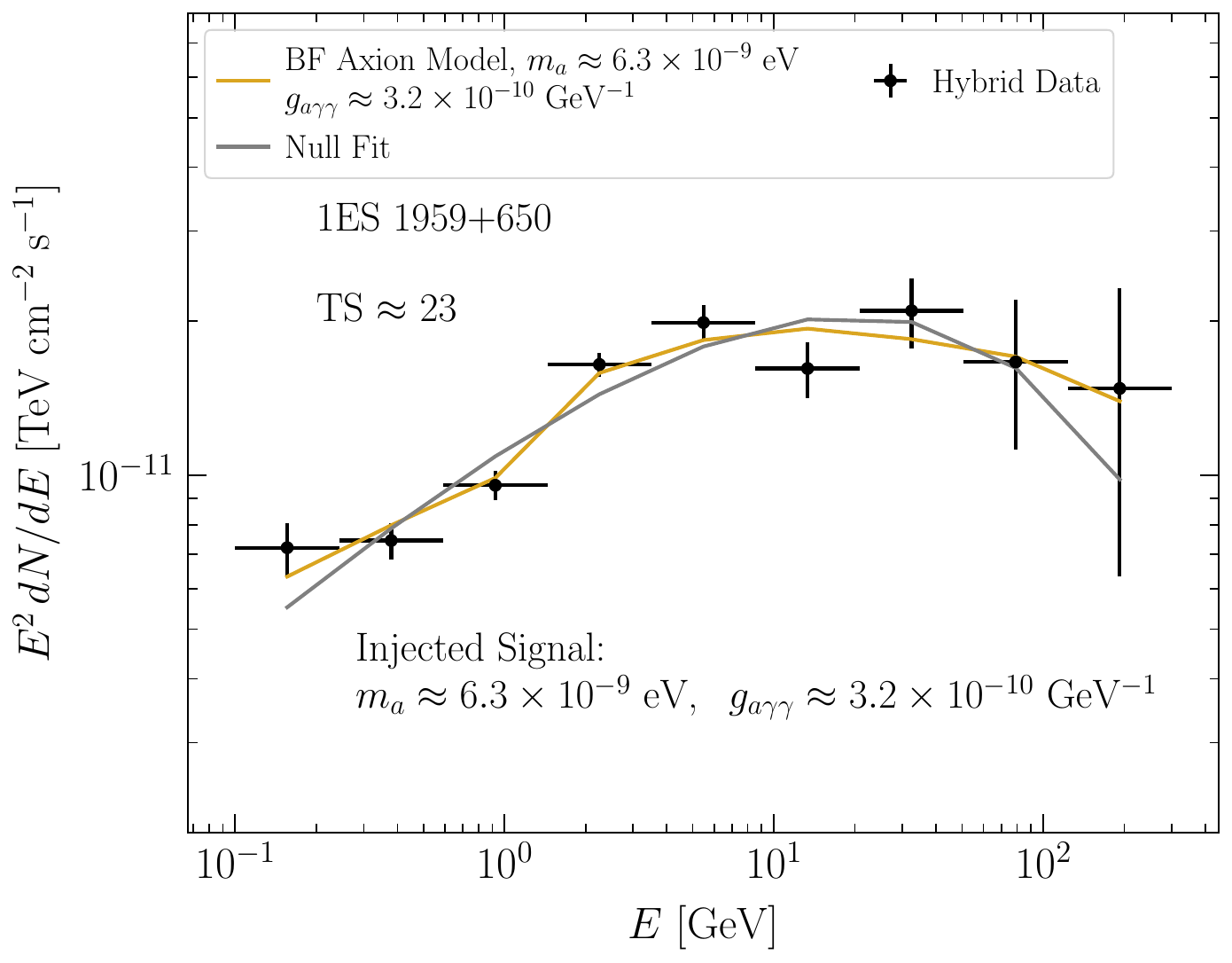}
    \caption{An illustration of the results of an injected signal test in blazar 1ES 1959+650, in which hybrid data containing the actual Fermi data altered by a synthetic axion-induced modulation (under Config. 4) is analyzed. The resulting signal, with the injected mass and coupling indicated, is recovered in our statistical analysis as the best-fit axion parameter combination, with the indicated discovery TS.}
    \label{fig:siginj}
\end{figure}

In this section, we examine the robustness of our analysis pipeline by performing a signal injection test, where we insert an axion-induced modulation into our fiducial data for blazar 1ES 1959+650 (here we assume the modulation follows the parameters given in Config. 4). We then apply our analysis to the hybrid data. As illustrated in Fig.~\ref{fig:siginj}, the modulation induces distinct hybrid data that highly favors the axion signal model with the appropriately recovered $(m_a, \gagg)$ combination reflecting the injected signal strength. Our analysis framework is thus capable of detecting high-significance axion-induced modulations that may be in our data, for a large enough signal.

\clearpage
\newpage

%

\end{document}